\documentclass[aps,pra,showpacs,twocolumn,groupedaddress]{revtex4}
\usepackage{amsmath}
\usepackage{amssymb}
\usepackage{graphics}
\usepackage[dvips]{graphicx}
\usepackage{graphicx}% Include figure files
\usepackage{color}

\newcommand \beq{\begin{eqnarray}}
\newcommand \eeq{\end{eqnarray}}
\def\simge{\mathrel{%
       \rlap{\raise 0.511ex \hbox{$>$}}{\lower 0.511ex \hbox{$\sim$}}}}
\def\simle{\mathrel{
       \rlap{\raise 0.511ex \hbox{$<$}}{\lower 0.511ex \hbox{$\sim$}}}}

\usepackage{graphicx,color}
\usepackage{amsmath,amssymb}
\usepackage{feynmf}
\usepackage{young}
\newcommand{\Slash}[1]{{\ooalign{\hfil/\hfil\crcr$#1$}}}
\newcommand{\tr}{{\rm tr}}

\newcommand{\Nc}{N_{\rm c}}
\newcommand{\Nf}{N_{\rm f}}

\newcommand{\lqcd}{\Lambda_{\rm QCD}}

\newcommand{\vp}{\vec{p}}
\newcommand{\vq}{\vec{q}}
\newcommand{\vk}{\vec{k}}

\newcommand{\la}{\langle}
\newcommand{\ra}{\rangle}

\newcommand{\calL}{\mathcal{L}}

\newcommand{\calS}{\mathcal{S}}  

\newcommand{\calA}{\mathcal{A}}

\newcommand{\calD}{\mathcal{D}}

\newcommand{\calP}{\mathcal{P}}

\newcommand{\rmd}{\mathrm{p}}
\newcommand{\rma}{\mathrm{a}}
\newcommand{\rmi}{\mathrm{i}}
\newcommand{\rme}{\mathrm{e}}

 \newcommand{\itDelta}{{\mathit \Delta}}
\newcommand{\bPi}{ \bar{\Pi}}

\begin{document}
\title{Color screening in cold quark matter }
\author{Toru Kojo and Gordon Baym}
\affiliation{Department of Physics, University of Illinois, 1110
  W. Green Street, Urbana, Illinois 61801, USA}  
\date{\today}

\begin{abstract}
We compute---at finite quark chemical potentials---the color screening of cold quark matter
at the one-loop level, comparing the normal, BCS-paired U(1)$_{ {\rm em} }$ (or Higgs)
phase 
and a singlet phase with 
color-singlet condensate near the Fermi surface.  
The latter phase is computed using the example of two-color
QCD with a color-singlet diquark condensate.
In contrast to the normal and Higgs phases,
neither electric nor magnetic screening masses 
appear in the singlet phase.
The absence of a magnetic mass, within a perturbative framework, is a consequence of
the proper treatment of gauge invariance.
While at large momenta 
the gluon self-energies approach  
those in the normal phase,
the medium contributions to the 
infrared region below a scale of the mass gap
are substantially suppressed.  Infrared gluons at low quark density in the singlet phase
appear  protected from medium effects,
unless the quark-gluon vertices
are significantly enhanced in the infrared.
\end{abstract}

\maketitle

%%%%%%%%%%%%%%%%%%%%%%%%%%%%%%%%%%%%%%
\section{Introduction}
%%%%%%%%%%%%%%%%%%%%%%%%%%%%%%%%%%%%%%

Many properties of degenerate quark matter at large quark densities
can be understood within a picture of weakly coupled  quarks and gluons
\cite{Collins:1974ky,Freedman:1976xs,Alford:2007xm}.  However, at intermediate quark chemical potentials, $\mu \sim \lqcd$,
where $\lqcd$ is the QCD scale parameter, strong coupling dynamics intrinsic to QCD dominate, until
strong screening by the medium sets in.
Our aim in this paper is to delineate screening effects of the medium on
the gluon dynamics in this
intermediate regime, a regime relevant for phenomenology, e.g., the physics of neutron stars. 

The effects of the medium on the gluon polarization, or self-energy, depend on
how quarks participate in the screening 
processes \cite{Rischke:2000qz,Iida:2001pg}. 
A condensate formed near the Fermi surface
affects the quark mass gap $\Delta$ as well as 
the effective quark-gluon coupling, indicating the need to  
determine screening effects and mass gaps self-consistently in finite-density quark matter. 

\begin{figure}[tb]
\vspace{-0.5cm}
\begin{center}
  \includegraphics[scale=.20]{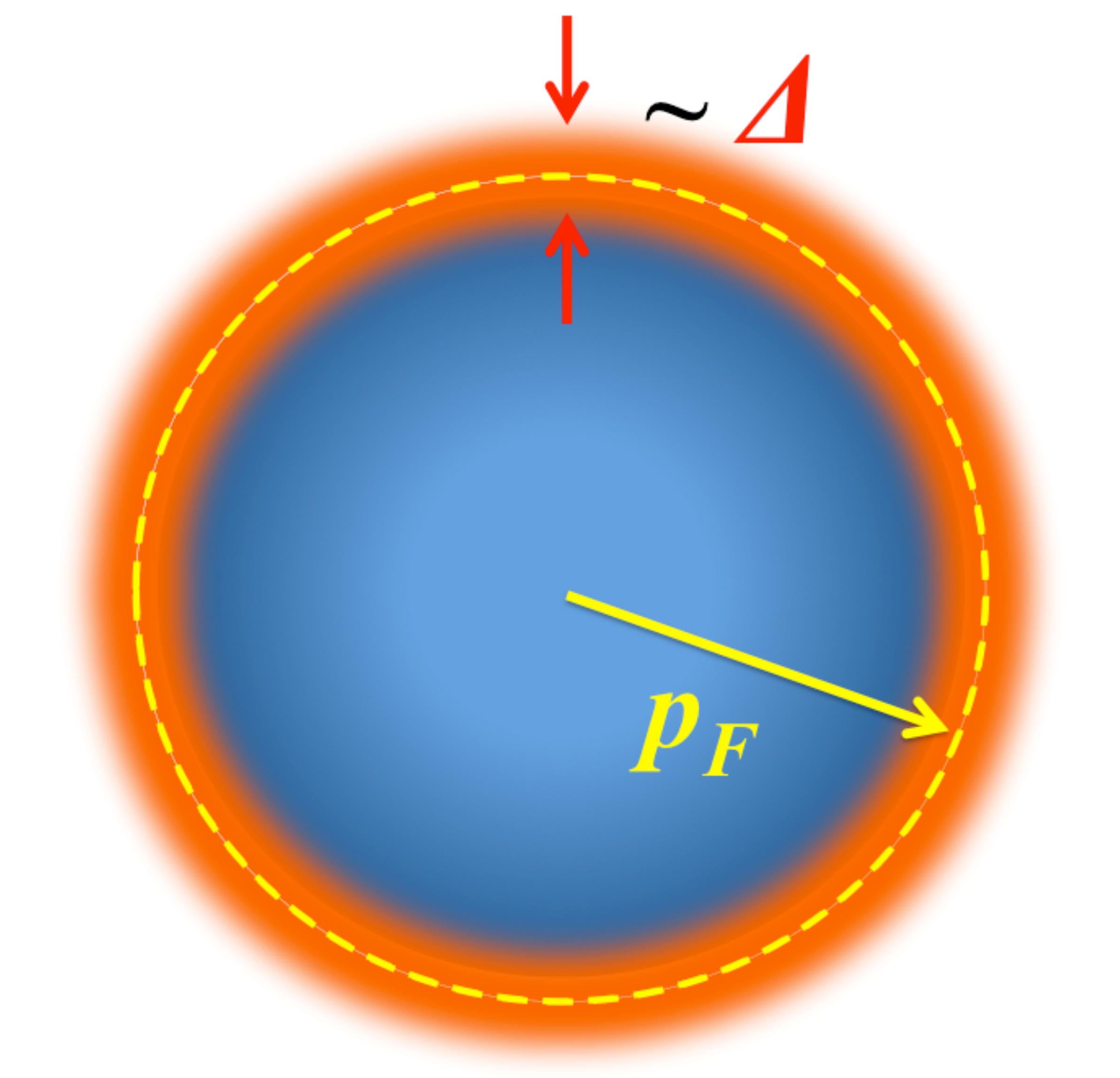} 
\end{center}
\vspace{-0.6cm}
\caption{A Fermi sea of quarks
condensed in a domain around the Fermi surface of thickness
$\sim \Delta$.
Soft gluons mainly perturb soft 
quarks close to the Fermi surface, which in turn interact back on the gluons.
Quarks with large mass gaps hardly react to perturbations 
from soft gluons with momenta smaller than $\sim \Delta$.
The exception is the BCS-condensed (or Higgs phase) in which
soft colored {\it phase} fluctuations of the condensate
strongly react, giving a Meissner mass.  
This feedback does not occur
if the condensate is a
color singlet, in which
soft gluons are protected from backreactions.}
\label{fig:Fermisurface}
\end{figure}

The importance of the gap on quark properties depends on the size of the domain over which quark
self-energies are modified (Fig.~\ref{fig:Fermisurface}).
In weak coupling with a small gap, the gap is relevant only in the very vicinity of the Fermi surface;
elsewhere quarks behave normally.  Consequently, the effects of 
the gap are important only for gluons of momenta smaller than the gap.
Therefore in weak coupling 
one does not need to take into account effects of the gap on the gluon self-energy, which arise 
only in a limited region of phase space.
Such a picture allows one to use 
in-medium gluon propagators computed in the hard-dense-loop limit,
which Son \cite{Son:1998uk} used to reliably estimate the color superconducting
gap for $\mu \gg \lqcd$.

The situation differs for strong coupling, $\alpha_s \sim 1$,
where gaps are $\sim \lqcd$, and a substantial fraction of quarks---
within a domain $|E_q -\mu| \sim \Delta$
behave differently from those in
the normal phase.
The number of soft gluons affected by such quarks
is no longer small, and thus the nature of the gluon sector
can be substantially different 
from that in normal quark matter.  Thus, extrapolation of the hard-dense-loop picture
is inconsistent and potentially misleading.

In this paper
we examine how a large
quark mass gap, $\sim \lqcd$, affects screening 
in different possible phases of dense matter, including the normal phase---a 
U(1)$_{ {\rm em} }$ BCS-paired, or ``Higgs phase"---and 
a phase with a quark color singlet condensate.  We investigate the gluon polarization in the one-loop approximation 
over wide regions of momenta and values of the gap, $\Delta$.
While one-loop calculations include nonperturbative effects related to
condensation, they basically
extrapolate the perturbative picture, and thus cannot allow one
to reach definitive conclusions, which may depend on further nonperturbative physics.
However they allow one to investigate
how screening effects differ
for various phases at the same order of loops, and furthermore, they
serve as a useful step
to sharpen questions
about nonperturbative gluon dynamics
at finite density.

In determining the gluon polarization function it is critical to include 
gauge invariance or, equivalently, the related conservation laws
\cite{Nambu:1960tm,Baym:1961zz} 
to avoid artificial contributions in the simplest one-loop calculations.
We discuss, in this context, the physics of a constant gap as 
well as one dependent on momentum, and we will see that different
artifacts arise in these two cases, requiring different resolutions;
in the former the artifact typically comes from
a gauge-variant regularization, while 
in the latter the problem arises from use of the bare---rather than renormalized---vertex.
As we show,  the artifacts, for $\Delta \sim \lqcd$,
are not negligible
and even affect qualitative interpretations,
especially in the magnetic sector.

We compare in this paper characteristic screening features
of the normal phase---a
U(1)$_{ {\rm em} }$ Higgs phase---and 
a color-singlet quark condensate, which
we call a ``singlet phase.''  In particular we study the realization of the phase
in two-color QCD as a color-singlet diquark
condensate near the Fermi surface.
The two-color QCD system is theoretically simpler than 
that with three colors,
and moreover
lattice studies are possible due to the
absence of the sign 
problem \cite{Nakamura:1984uz,Kogut:2001na,Cotter:2012mb}, 
making this a well-suited system for illustrative purposes.

Let us briefly overview the characteristic features
of the gluon polarization function in these three phases.
In normal quark matter, quarks are gapless near the Fermi surface.
The one-loop results for in-medium electric and magnetic 
screening masses, $m_E$ and $m_M$,   for
for a number of flavors $\Nf =2$, are
\beq
\frac{\, m^2_E(k) \,}{\, g_s^2 (k) \,} \bigg|_{k_0=0,\vk\rightarrow 0}
= \frac{ \Nf \mu^2}{\pi^2} 
\simeq 0.2 \, \mu^2 \nonumber\\
({\rm normal~phase}),
\eeq
for gluons with small momenta $\vk$.
(Here we divide $m_E^2$ by $g_s^2$, 
because we are interested in
comparing the vacuum gluon polarization 
and in-medium effects, both of which have
an overall factor, $\sim g_s^2$.)
Electric screening is well dominated by 
particle-hole excitations near the Fermi surface,
and the factor $\mu^2$ reflects 
the fact that their phase space is
proportional to the area of the Fermi surface.
Taking this expression at face value, we see that
at low density, $\mu \sim (1.5-2.0) \lqcd \sim (0.3-0.4)$ GeV, 
\beq
m^2_E /g_s^2 \sim (0.45-0.8) \, \lqcd^2 
~~({\rm normal~quark~matter});
\eeq
this rough estimate implies that electric screening is 
sizable even for small quark Fermi seas,
and medium contributions may dominate vacuum screening effects.

On the other hand, there is no screening in the magnetic sector
at zero frequency; rather gluons undergo
Landau damping at finite frequency \cite{Baym:1990uj},
\beq
m_M^2(k) \simeq -\rmi \, \frac{\, k_0 \,}{|\vk|} \, m_E^2 
~~~~({\rm normal~phase}),
\eeq
for $k_0\ll |\vk|$.
The absence of a magnetic screening (or Meissner) mass
is a consequence of exact cancellations between the paramagnetic
particle-hole contributions and the diamagnetic
particle-antiparticle contributions.

We assume, in discussing a U(1)$_{ {\rm em}}$ Higgs phase, that the system is charge neutral,  that the
gap  $\sim \lqcd$ is generated by attractive quark interactions, and that the coupling constant is the QCD coupling $g_s$
instead of the electromagnetic coupling constant.  This phase has an
 electric  as well as magnetic mass.
 At one-loop, the screening masses are
\beq
m^2_E/g_s^2 \sim \mu^2 \,,
~~~~~~~
m^2_M/g_s^2 \sim \mu^2 \,.
~~~~({\rm Higgs~phase})
\eeq
Both the electric and magnetic masses are of the scale $\mu$.
On the other hand, the size of
domain in momentum space that differs from the normal phase is
$\sim\Delta$.
Beyond $k\sim \Delta$, both screening masses approach those in the normal phase.
In contrast to the situation in weak coupling, $\mu \gg \lqcd$, the dominant
magnetic interactions are characterized by a
Meissner mass instead of Landau damping; the latter
occurs only at $|\vk|$ beyond $\Delta$ due to
the lack of phase space for decays.

Finally we consider a phase with a color-singlet condensate.  In the absence of color pairing,
the magnetic mass should vanish.  In addition, as we shall see, screening in the gluon electric sector is also absent
as a consequence of the quark mass gap, 
together with the vanishing at long wavelengths of the matrix element between
the color current and quark particle-hole excitations.
As we show below,
the masses in the gluon polarization function behave at small $\vk$ like:
\beq
m^2_E(k)/g_s^2 \sim \vk^2 
\times \frac{\mu^2}{\, \Delta^2 \,} + \cdots  \nonumber\\
m^2_M(k) \sim 0 
~~~~({\rm singlet~phase}).
\eeq
Thus gluons in the infrared limit 
are protected from medium effects; 
instead,
screening effects renormalize the effective color charges, as in the magnetic sector of the normal phase.
In the magnetic sector, Landau damping, as in the Higgs phase,
is operative only for $|\vk|$ beyond $\sim \Delta$.  Due to suppression of electric and magnetic
screening effects for the gluons in medium that produce the color-singlet condensate, one can expect much larger quark mass gaps
than those obtained with a hard-dense-loop gluon propagator.  This regime is somewhat similar
to that in quark matter 
at large $\Nc$ \cite{McLerran:2007qj},
in which quark-loop effects can be ignored
and gluons remain strongly interacting.

We mention candidates of phases in 
three-color QCD for which our results for the singlet phase can
provide insights.
The first is the inhomogeneous chiral condensate phase
\cite{Migdal:1978az,Deryagin:1992rw}
which recently attracted renewed attention in 
quark matter at strong 
coupling \cite{Nickel:2009wj,Kojo:2009ha}.
The condensation is mainly created here by 
color-singlet particle-hole
pairs of nonzero total momentum. 
Some (but not all) inhomogeneous phases
open a mass gap near the Fermi surface.  This
situation has a resemblance to 
the singlet phase in this paper.

Another interesting example is 
the two-flavor color superconductor with residual
unbroken color SU$_{\rm c}(2)$ symmetry,
 investigated by in detail by Rischke \cite{Rischke:2000qz}.
Although the condensate is not a color singlet, 
aspects of the physics in that case are similar to those in the singlet phase:
SU$_{ {\rm c} }(2)$ gluons do not directly couple to
the diquark condensate and thus do not acquire
a Meissner mass, 
and at the same time gapped quarks
do not generate an electric mass either. Conceptually the present study has several overlaps
with the proposed asymptotic deconfinement
scenario at high density of Rischke, Son, and Stephanov \cite{Rischke:2000cn}.

This paper is organized as follows.
In Sec.~\ref{quarkbases} we fix our conventions
and summarize standard techniques,
such the Nambu-Gor'kov basis, 
particle-antiparticle projection operators, etc.
In Sec.~\ref{generalremarks}
we remark generally on aspects of 
polarization functions,
such as vacuum subtraction,
constraints from conservation laws,
and the minimally improved vertex.
In Sec.~\ref{1loop} we present 
the simplest one-loop calculations, and
in Sec.~\ref{improvement}
we 
take into account gauge invariance, thus
erasing artificial contributions and correcting the simple one-loop result.  Then in
Sec.~\ref{numerical} we show numerical results for the gluon polarization function 
for various momenta and gaps.  The final section, Sec. \ref{summary}, is devoted to a summary and discussion of possible phenomenological applications, and a confrontation of our results
with the lattice data for Landau gauge propagators.

   In this paper we consider only zero temperature
and the two-flavor limit with equal
masses for the $u$ and $d$ quarks.  We work in Euclidean coordinates,
$p_\mu = p^\mu = (p_0, \vp\,)$, with $\gamma$ matrices
satisfying
$\{\gamma_\mu, \gamma_\nu \}=2\delta_{\mu \nu}$,
$\gamma_\mu = \gamma_\mu^\dag$, 
and $\gamma_5 = \gamma_5^\dag$.
The partition function and  action are related 
as $Z = \int \rme^{-\calS} = \int \rme^{-\int \calL}$ 
and the fermionic part of the Lagrangian is given by
$\calL_{ {\rm f} } = \bar{\psi} \left( \Slash{D} +m \right) \psi$
with $\psi^T=(u,d)$ and $D_\mu = \partial_\mu + \rmi A_\mu^a T_a$,
where $T_a =\sigma_a/2$ is the SU(2) color matrix.  We absorb
the gauge coupling constant $g_s$ 
into definition of the gluon field $A_\mu^a$, and take the gauge
action $\tr\, G_{\mu \nu}^2/2g_s^2$,
where $G_{\mu \nu}$ is the usual field-strength tensor.
For flavor we use the Pauli matrices $\tau_f$.
We denote the energy of a normal particle by
$E_q = \sqrt{\vq^2 + m^2}$, and use $\epsilon(q)$ for 
the excitation energy of a quasiparticle (and
quasiantiparticle)
We also use the shorthand $\int_x \equiv \int {\rm d}^4x$
and $\int_k \equiv \int {\rm d}^4k /(2\pi)^4$.

%%%%%%%%%%%%%%%%%%%%%%%%%%%%%%%%%%%%%%
\section{Two-color QCD}
\label{quarkbases}
%%%%%%%%%%%%%%%%%%%%%%%%%%%%%%%%%%%%%%

We consider here diquark condensation in two-color QCD with  two flavors.
The simplest condensate, a color and flavor singlet, with quantum number $J^{PC}=0^{++}$,
breaks U(1)$_B$ symmetry, but keeps color and flavor symmetry unbroken.
In this channel, both the color electric and magnetic  interactions are attractive, 
which implies that this condensate is the most favored.
Because the color and flavor wave functions are totally  anti-symmetric, 
the diquark condensate takes the form
\beq
d =
- \epsilon_{cc'} \epsilon_{ff'} 
\left\la \left(\bar{\psi}_{C}\right)_c^f \gamma_5 \, \psi_{c'}^{f'} \right\ra 
= \left\la \, \bar{\psi}_{C} \tau_2 \sigma_2 \gamma_5 \psi
  \,\right\ra ,
\eeq
where we use $(\tau_2)_{fg}= \rmi \epsilon_{fg}$, etc. where
$\epsilon$ is the antisymmetric tensor in two indices;
the factor $\gamma_5$ gives positive parity.
To describe the U(1)$_{ {\rm em} }$ Higgs phase,
we simply replace $\tau_2 \sigma_2 $ by unity.  We construct the
quark propagator
with a self-energy that yields
the desired condensate structure. 

We carry out all computations in the Nambu-Gor'kov basis 
(see Ref. \cite{Pisarski:1999av} for useful details),
with the spinor 
$\Psi (x) \equiv (\psi, \psi_C)^T/\sqrt{2}$,
where
$\psi_C(x) \equiv C \bar{\psi}^T (x)$ \footnote{In Euclidean space $C=\gamma_2 \gamma_0$
satisfies the usual relations, $C=-C^{-1}=-C^T=-C^\dag$, and $C\gamma_\mu^T C^{-1} = -\gamma_\mu$.}.
In this basis, the quark kinetic and mass terms are
\beq
\calL_{ {\rm kin,mass} }
&=& \bar{\psi} \left[\, 
\Slash{\partial} -\mu \gamma_0 + m \, \right] \psi \nonumber\\
&=&  \bar{\Psi}
\left[\begin{matrix}
~ \Slash{\partial} -\mu \gamma_0 + m  ~&~ 0 ~\\
~0 ~&~ \Slash{\partial} + \mu \gamma_0 + m ~ 
\end{matrix}
\right] \Psi,
\eeq
and the quark-gluon vertex is
\beq
\calL_{\rm int}
= \rmi  \, \psi \gamma_\mu T_a A^a_\mu  \psi
= \rmi \bar{\Psi}\, \Gamma_\mu^a  A_\mu^a \Psi\,,
\eeq
where
\beq
\Gamma_\mu^a \equiv \gamma_\mu R_a\,,
~~~~R_a \equiv
\left[\begin{matrix}
~~ T_a  ~& ~0 ~\\
~0~ & -T_a^T~  \,
\end{matrix}\right]\,.
\label{R}
\eeq
The transpose of the color matrix appears
because the color transformation, 
$\psi \rightarrow \rme^{\rmi \theta_aT_a} \psi$,
gives a rotation,
$\psi_C \rightarrow \rme^{-\rmi \theta_a T^T_a} \psi_C$.
For the U(1)$_{ {\rm em} }$ Higgs phase,
we replace $T_a$ by unity.

  To generate a diquark condensation, we include the spatially dependent
quark self-energy in the quark propagator,  
\beq
\Sigma(q) = 
\left[\begin{matrix}
~0~&  {\bf \bar{\Delta} }(q) ~ \\
~ {\bf \Delta}(q) & ~0~ \,
\end{matrix}
\right].
\label{Sigma}
\eeq
where the energy gap is given by
\beq
{\bf \Delta}(q)
= \tau_2 \sigma_2 \gamma_5
\left(\, \Delta_\rmd(q) \Lambda_\rmd(q) 
+ \Delta_\rma (q) \Lambda_\rma (q) \, \right)\,,
\eeq
and
${\bf \bar{\Delta} }(q)
=\gamma_0  {\bf \Delta}^\dag (q) \gamma_0$.
The particle gap $\Delta_\rmd$
and the antiparticle gap  $\Delta_\rma$ depend differently on condensates
formed near the Fermi surface.  The particle and antiparticle projection operators
are defined by
\beq
\Lambda_{\rmd, \rma} (q)
= \gamma_0
\frac{\, E_q \gamma_0
\pm \left(m -\rmi \vec{\gamma} \cdot \vq \right) \,}{2E_q}.
\eeq
and have the properties,
\beq
\Lambda_{\rmd} + \Lambda_\rma = 1\,,~~
\Lambda_{\rmd, \rma}^2 = \Lambda_{\rmd,\rma}\,,~~
\Lambda_\rmd \Lambda_\rma =0\,, 
~~ \Lambda_{\rmd, \rma}^\dag = \Lambda_{\rmd, \rma}. \nonumber\\
\eeq
We further introduce the projection operators for
charge-conjugated fields,
\beq
\Lambda^C_{\rmd,\rma} = \Lambda_{\rma,\rmd},
\eeq
which obey the useful relations
\beq
\gamma_0 \Lambda_{\rmd, \rma} \gamma_0
= \gamma_5 \Lambda^C_{\rmd, \rma} \gamma_5 \,.
\eeq

We define the quark propagator
\beq
\calS
=
i\left[\begin{matrix}
~ \la \psi \bar{\psi} \ra ~&~ \la \psi \bar{\psi}_C \ra  ~ \\
~ \la \psi_C \bar{\psi} \ra~ & ~  \la \psi_C \bar{\psi}_C \ra ~
\end{matrix}
\right].
\label{form}
\eeq  
With the self-energy (\ref{Sigma}), 
\beq
\calS^{-1}(q)
=
\left[\begin{matrix}
~ -\rmi \Slash{q} -\mu \gamma_0 + m~&  {\bf \bar{\Delta} }(q) ~ \\
~ {\bf \Delta}(q) & ~-\rmi \Slash{q} +\mu \gamma_0 + m~~
\end{matrix}
\right].
\eeq
The propagator yields quasiparticles of energy $\pm \epsilon _{\rmd,\rma}(q)$ where
\beq
\epsilon_\rmd (q) 
= \sqrt{\, (E_q-\mu)^2 + |\Delta_\rmd (q)|^2 \,} \,,\nonumber\\
\epsilon_\rma (q) 
= \sqrt{\, (E_q+\mu)^2 + |\Delta_\rma (q)|^2 \,} \,.
\label{S}
\eeq
The nonanomalous and anomalous parts of the propagator are  
\beq
\calS(q)
=
\left[\begin{matrix}
~ \calS^D_{11} (q) ~&~ \tau_2 \sigma_2 \, \calS^D_{12} (q)  ~ \\
~ \tau_2 \sigma_2 \, \calS^D_{21} (q)~ & ~  \calS^D_{22} (q) ~
\end{matrix}
\right],
\label{form}
\eeq
where the $\calS^D$ contain the Dirac structures.  The normal parts are, explicitly,
\beq
\calS_{11}^D(q)
&=& - \left[
\frac{\,  \left| u_\rmd (q) \right|^2 \,}{\, \rmi q_0 - \epsilon_\rmd(q) \,} 
+ \frac{\, \left| v_\rmd (q) \right|^2 \,}{\, \rmi q_0 + \epsilon_\rmd(q) \,} 
\right]
\Lambda_\rmd \gamma_0 \nonumber\\
&&-  \left[
\frac{\, \left| v_\rma (q) \right|^2 \,}{\, \rmi q_0 - \epsilon_\rma (q) \,} 
+ \frac{\, \left| u_\rma (q) \right|^2 \,}{\, \rmi q_0 + \epsilon_\rma (q) \,} 
\right]
\Lambda_\rma \gamma_0 \,,
\label{normalpropagator1}
\eeq
and
\beq
\calS_{22}^D (q)
&=& - \left[
\frac{\, \left| v_\rmd (q) \right|^2 \,}{\, \rmi q_0 - \epsilon_\rmd(q) \,} 
+ \frac{\, \left| u_\rmd (q) \right|^2 \,}{\, \rmi q_0 + \epsilon_\rmd(q) \,} 
\right]
\Lambda^C_\rmd \gamma_0 \nonumber\\
&&- \left[
\frac{\, \left| u_\rma (q) \right|^2 \,}{\, \rmi q_0 - \epsilon_\rma (q) \,} 
+ \frac{\, \left| v_\rma (q) \right|^2 \,}{\, \rmi q_0 + \epsilon_\rma (q) \,} 
\right]
\Lambda^C_\rma \gamma_0,
\label{normalpropagator2}
\eeq
while the anomalous parts are
\beq
\calS^D_{12} (q)
&=& - \left[
\frac{\, u^*_\rmd v^*_\rmd (q)\,}{\, \rmi q_0 - \epsilon_\rmd (q) \,} 
- \frac{\,u^*_\rmd v^*_\rmd (q) \,}{\, \rmi q_0 + \epsilon_\rmd (q) \,} \right]
\Lambda_\rmd \gamma_5  \nonumber\\
&&-\left[
\frac{\, u^*_\rma v^*_\rma (q) \,}{\, \rmi q_0 - \epsilon_\rma (q) \,} 
- \frac{\,u^*_\rma v^*_\rma (q)\,}{\, \rmi q_0 + \epsilon_\rma (q) \,} \right]
\Lambda_\rma \gamma_5,
\label{anomalouspropagator1}
\eeq
and
\beq
\calS^D_{21} (q)
&=& \left[
\frac{\, u_\rmd v_\rmd (q)\,}{\, \rmi q_0 - \epsilon_\rmd (q) \,} 
- \frac{\,u_\rmd v_\rmd (q) \,}{\, \rmi q_0 + \epsilon_\rmd (q) \,} \right]
\Lambda^C_\rmd \gamma_5  \nonumber\\
&&+ \left[
\frac{\, u_\rma v_\rma (q) \,}{\, \rmi q_0 - \epsilon_\rma (q) \,} 
- \frac{\,u_\rma v_\rma (q) \,}{\, \rmi q_0 + \epsilon_\rma (q) \,} \right]
\Lambda^C_\rma \gamma_5 .
\label{anomalouspropagator2}
\eeq
The coherence factors $u$ and $v$ obey
\beq
\left| u_{\rmd, \rma} (q) \right|^2
= \frac{\, 1 \,}{ 2 }
\left( 1 + \frac{\, E_q \mp \mu \,}{\epsilon_{\rmd, \rma} (q) } \right) \,, \nonumber\\
\left| v_{\rmd, \rma} (q) \right|^2
= \frac{\, 1 \,}{ 2 }
\left( 1 - \frac{\, E_q \mp \mu \,}{\epsilon_{\rmd, \rma} (q) } \right);
\eeq
then $|u_\rmd|^2 + | v_\rmd |^2 
= |u_\rma |^2 + | v_\rma |^2 =1$,
and
\beq
u_{\rmd}  v_\rmd(q) 
= \frac{\, \Delta_\rmd (q) \,}{\, 2\epsilon_\rmd (q)\,} \,, 
~~~~
u_{\rma} v_\rma (q) 
= \frac{\, \Delta_\rma (q) \,}{\, 2\epsilon_\rma (q)\,} \,.
\eeq
The gap functions and resulting condensates as well as the coherence factors
can have complex phases associated with the violation of U(1)$_B$ symmetry.
For simplicity we chose the phases so that $u$ and $v$ are real and positive.

%%%%%%%%%%%%%%%%%%%%%%%%%%%%%%%%%%%%%%
\section{Gluon polarizations}
\label{generalremarks}
%%%%%%%%%%%%%%%%%%%%%%%%%%%%%%%%%%%%%%

In this section we remark on the structure of the polarization functions, vacuum subtraction and renormalization,
and the constraints from conservation laws, as well as improving the quark-gluon vertex.

%%%%%%%%%%%%%%%%%%%%%%%%%%%%%%%%%%%%%%
\subsection{Structures of the gluon polarization functions}
%%%%%%%%%%%%%%%%%%%%%%%%%%%%%%%%%%%%%%

At finite baryon density, the gluon propagator must satisfy rotational symmetry in space,
and thus it can be written generally 
in terms of  electric, magnetic, and longitudinal components as
\beq
D_{\mu \nu} (k)
= P^E_{\mu \nu} D_E(k) + P^M_{\mu \nu} D_M(k)
+ D^L_{\mu \nu} (k)\,, 
\eeq
where
\beq
 D_{E,M} (k) = \frac{g_s^2}{\, k^2 + {\bf \Pi}_{E,M} \,},
\eeq
and $k_\mu D^L_{\mu \nu}\neq 0$.
The projection operators, %
\beq
P^M_{ij} &=& \delta_{ij} - \frac{\, k_i k_j \,}{\vk^2}\,,\nonumber\\
P^M_{00} &=& P^M_{0i} =P^M_{i0}=0\,,\nonumber\\
P^E_{\mu \nu} &=& g_{\mu \nu}-\frac{\, k_\mu k_\nu \,}{k^2} - P_{\mu \nu}^M\,,
\eeq
satisfy the transversality condition,
$k_\mu P^{M,E}_{\mu \nu} = 0$, as well as 
$P^{E,M}_{\mu \alpha} P^{E,M}_{\alpha \nu} = P^{E,M}_{\mu \nu}$
and $P^E_{\mu \alpha} P^M_{\alpha \nu}=0$.
The function $D_{\mu \nu}^L$, which depends on the gauge fixing,  
can be anisotropic.
The polarization functions ${\bf \Pi}_{E,M}$ include antiscreening effects from gluon loops
as well as screening effects from quarks.  

We restrict ourselves here to calculating the gluon polarization at the one-loop level, 
which contains  
correlations of the quark color currents,
$j_\mu^a = \bar{\psi}\gamma_\mu t_a \psi
=\bar{\Psi} \Gamma_\mu^a \Psi$, but does not reflect the full non-Abelian structure.
At one loop,
\beq
\Pi_{\mu \nu}^{ab}(k)
&\equiv&  \int_x \rme^{\rmi kx}  
\left\la  j_\mu^a(x) j_\nu^b (0) \right\ra  \nonumber\\
&=& P^M_{\mu \nu} \Pi^{ab}_M(k) + P^E_{\mu \nu} \Pi^{ab}_E(k) 
+ \Pi_{\mu \nu}^{L,ab} (k) \,,
\eeq
where $a,b$ are quark color indices.  
In general $\Pi_{\mu \nu}^{L,ab} (k)$ 
can be anisotropic (and not necessarily simply of the form
 $\sim k_\mu k_\nu \Pi_L$.
At the one-loop level one has
$k_\mu \Pi^{ab}_{\mu \nu} = k_\mu \Pi^{L,ab}_{\mu \nu}  =0$ 
(see the discussions around Eq.~(\ref{loops})), but not beyond one loop,
because the quark color current is not separately conserved \footnote{
In the non-Abelian case $(D_\mu j_\mu)^a =0$
instead of $\partial_\mu j^a_\mu=0$.
On the other hand, 
the conserved color current
associated with global color symmetry is given by
$J_\mu^a=j_\mu^a + (j_{{\rm g, gh} })_{\mu}^a$
where $j_{{\rm g, gh} }$ contains
gluons and ghosts.}.
In the following, when $\Pi_{00}^L=0$, 
as in a properly gauge-invariant treatment,
$\Pi_E$ and $\Pi_M$ have the structures
\beq
\Pi_E(k) &=& \frac{k^2}{\, \vk^2 \,} \Pi_{00} (k) 
~~~[{\rm for}~ \Pi^L_{00} (k) =0 ]\,, \nonumber\\
\Pi_M(k) &=& \frac{1}{\,2\,} P_{\mu \nu}^M \Pi_{\mu \nu}^{ab} (k) \,,
\label{projection}
\eeq
where $\Pi_{00}$ is the 00 self-energy in the radiation gauge. 
We emphasize that application of projection operators
does not automatically guarantee that we
extract the physical contributions.
In fact, the artificial contributions can 
(as we will see in Sec.\ref{improvement}) appear in the
$g_{\mu \nu}$ component.
The physical $\Pi_E$ and $\Pi_M$ can be determined 
only after gauge-variant artifacts are identified and removed.

%%%%%%%%%%%%%%%%%%%%%%%%%%%%%%%%%%%%%%
\subsection{Vacuum subtraction and renormalization}
\label{sec:vacsubtraction}
%%%%%%%%%%%%%%%%%%%%%%%%%%%%%%%%%%%%%%

The polarization functions at finite density
in general contain particle-hole
and particle-antiparticle contributions.
As at zero density,
the particle-antiparticle contributions contain ultraviolet divergences which require renormalization.
We consider here renormalization by subtraction of appropriately constructed counterterms.
Once the vacuum is correctly renormalized there are no further divergences at finite density.
The renormalized $\Pi^R_{ {\rm vac} }$
and bare self-energies $\Pi_{ {\rm vac} }$ in vacuum
are related, with indices temporarily omitted, as 
\beq
\Pi^R_{ {\rm vac} } (k;\lambda_R)
= \Pi_{ {\rm vac} } (k) + \delta_{ {\rm c} } \Pi_{ {\rm vac} } (k;\lambda_R) \,,
\eeq
where $\lambda_R$ is the momentum scale at which one renormalizes the vacuum terms,
and $\delta_{ {\rm c} } \Pi_{ {\rm vac} } (k;\lambda_R)$
is the counterterm which (i) removes the UV divergence,
(ii) forces $\Pi^R_{ {\rm vac} } (k;\lambda_R)$
to be the experimental value at $k^2=\lambda_R^2$,
and (iii) restores symmetries
that can be artificially violated by the 
UV regularization scheme \footnote{The point here is that the regularized expression has divergent and finite terms.
Gauge-variant artifacts, which are hidden in the finite terms, must be eliminated by counterterms, which however contain divergent and finite pieces as well as
gauge-variant pieces if the regularization
is gauge-variant.}.
Using the counterterm defined in vacuum,
the renormalized self-energy at finite density is
\beq
\Pi^R (k;\lambda_R)
&=& \Pi  (k) + \delta_{ {\rm c} } \Pi_{ {\rm vac} } (k;\lambda_R) \nonumber\\
&=& \Pi^R_{ {\rm vac} } (k;\lambda_R)
+ \itDelta \Pi(k),
\label{reno}
\eeq
where
\beq
\itDelta \Pi(k) \equiv  \Pi (k) - \Pi_{ {\rm vac} } (k) \,.
\label{reno1}
\eeq
is a target of our computations.

Finally, we note that the problem
in applying the renormalized expression (\ref{reno})
to QCD computations for small $k$ is that
the vacuum expression at small $k$---which 
is considerably affected by non-perturbative effects---
is not precisely known.
Therefore we simply model $\Pi^R_{ {\rm vac} }(k;\lambda_R)$
and $\Pi_{ {\rm vac} }(k)$ with
the usual one-loop result, 
replacing current quark masses 
with the constituent quark masses, $M_\chi \sim 300$ MeV.
This treatment introduces additional ambiguities to
our estimates.

%%%%%%%%%%%%%%%%%%%%%%%%%%%%%%%%%%%%%%
\subsection{Gauge invariance
or the transversality condition}
\label{trans}
%%%%%%%%%%%%%%%%%%%%%%%%%%%%%%%%%%%%%%

In the simplest one-loop computations with the bare vertex,
two types of artificial contributions appear,
depending on whether the gap is constant or
momentum dependent.
Without the removal of these artifacts,
we would find, for instance, a nonzero color magnetic screening mass
even without symmetry breaking in color.

The first type of artifact is related to the regularization scheme.
In the usual loop computations at finite density,
we first pick up residues from the $q_0$ integration
and then integrate over spatial momenta $|\vq\,|$.
For this treatment to be unambiguous,
$|\vq\,|$ must be cut off at some UV scale $\Lambda$,
otherwise the residues with $|\vq\,| \rightarrow \infty$
may lie outside of the circle we draw
in the complex $q_0$ plane to pick up residues. 
But the introduction of the cutoff violates
particle conservation and, as a result, gauge invariance,
yielding regularization-dependent artifacts.
Such contributions must be removed using gauge-variant 
counterterms to restore gauge invariance in the final expression.

The second type is more physical.
The self-energy term, $\sim \int_{x,y} \bar{\psi}_C(x) \Delta(x-y) \psi(y)$,
in coordinate space is nonlocal and breaks local color-gauge invariance, since under a gauge transformation
it transforms as
\beq
\bar{\psi}_C(x) \Delta(x-y) \psi(y) & \nonumber \\
\rightarrow  \bar{\psi}_C (x) 
\Delta(x-y) & \rme^{-\rmi T_a \theta_a(x) } 
\rme^{\rmi T_a \theta_a(y)  } \psi(y) \,,
\eeq
where we used 
$\rme^{ \rmi T_a^T \theta_a } \tau_2 = \tau_2 \rme^{ - \rmi T_a \theta_a } $.
This self-energy term is invariant for a constant gap,
$\Delta(x-y) = \Delta \delta^D (x-y)$,
but not for one with momentum dependence.
In the former case, as we show, using the bare vertex in the loop
is sufficient to maintain the transversality condition.
In the latter case, however, it is essential 
to use the improved vertices to eliminate the gauge-variant
components in the approximate quark propagators.
As clarified by Nambu for the BCS theory \cite{Nambu:1960tm},
the Ward-Takahashi identity 
can be used to constrain the form of the longitudinal 
vertex through the quark propagators,
and such a vertex can cure the transversality condition.   Equivalently, one needs to derive the 
self-energy self-consistently via a $\Phi$-derivable approximation \cite{Baym:1961zz}.  \

\begin{figure}[tb]
\vspace{0.0cm}
\begin{center}
\scalebox{0.4}[0.4] {
  \includegraphics[scale=.43]{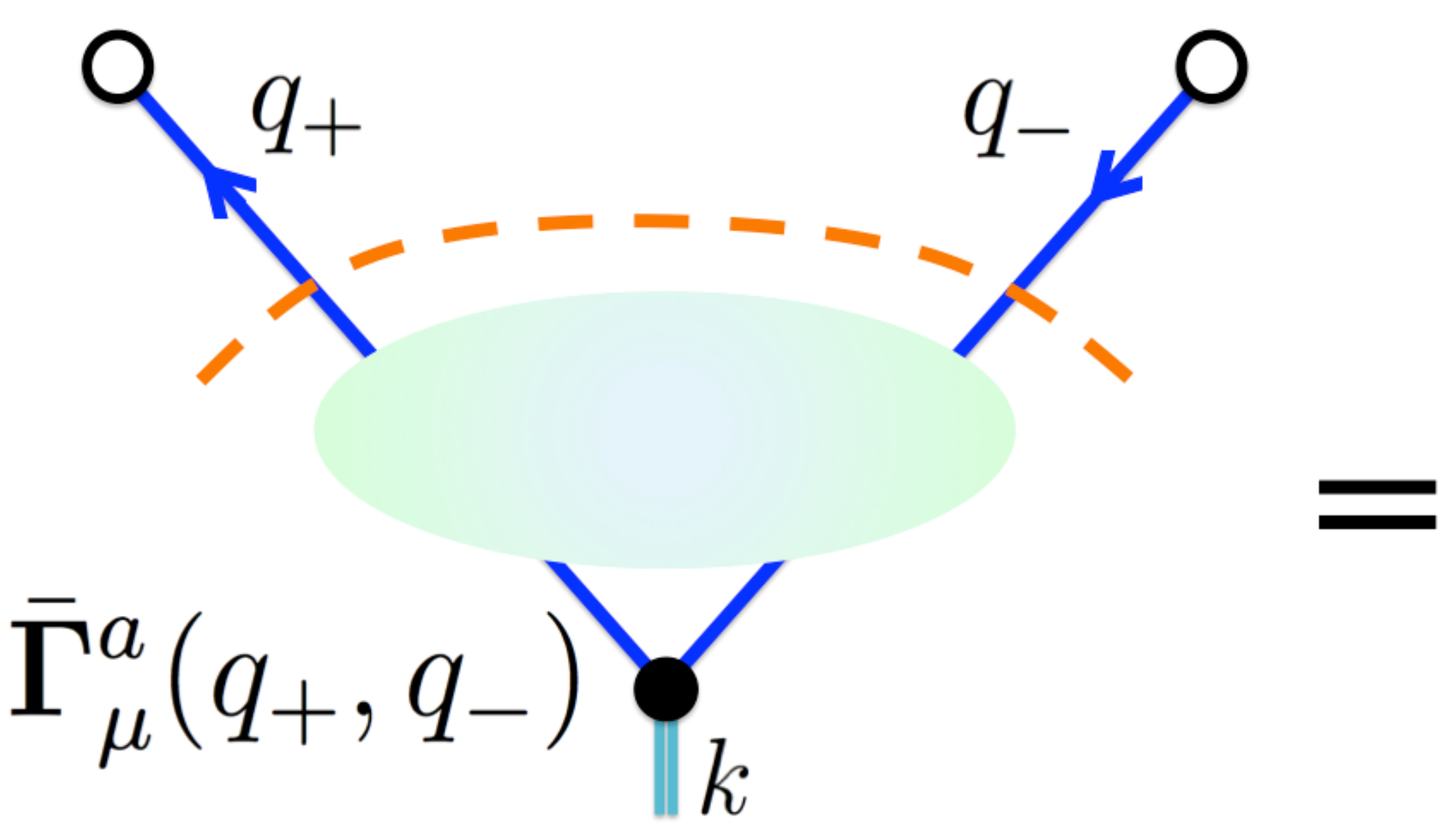} 
} \hspace{0.0cm}
\scalebox{0.4}[0.4] {
  \includegraphics[scale=.43]{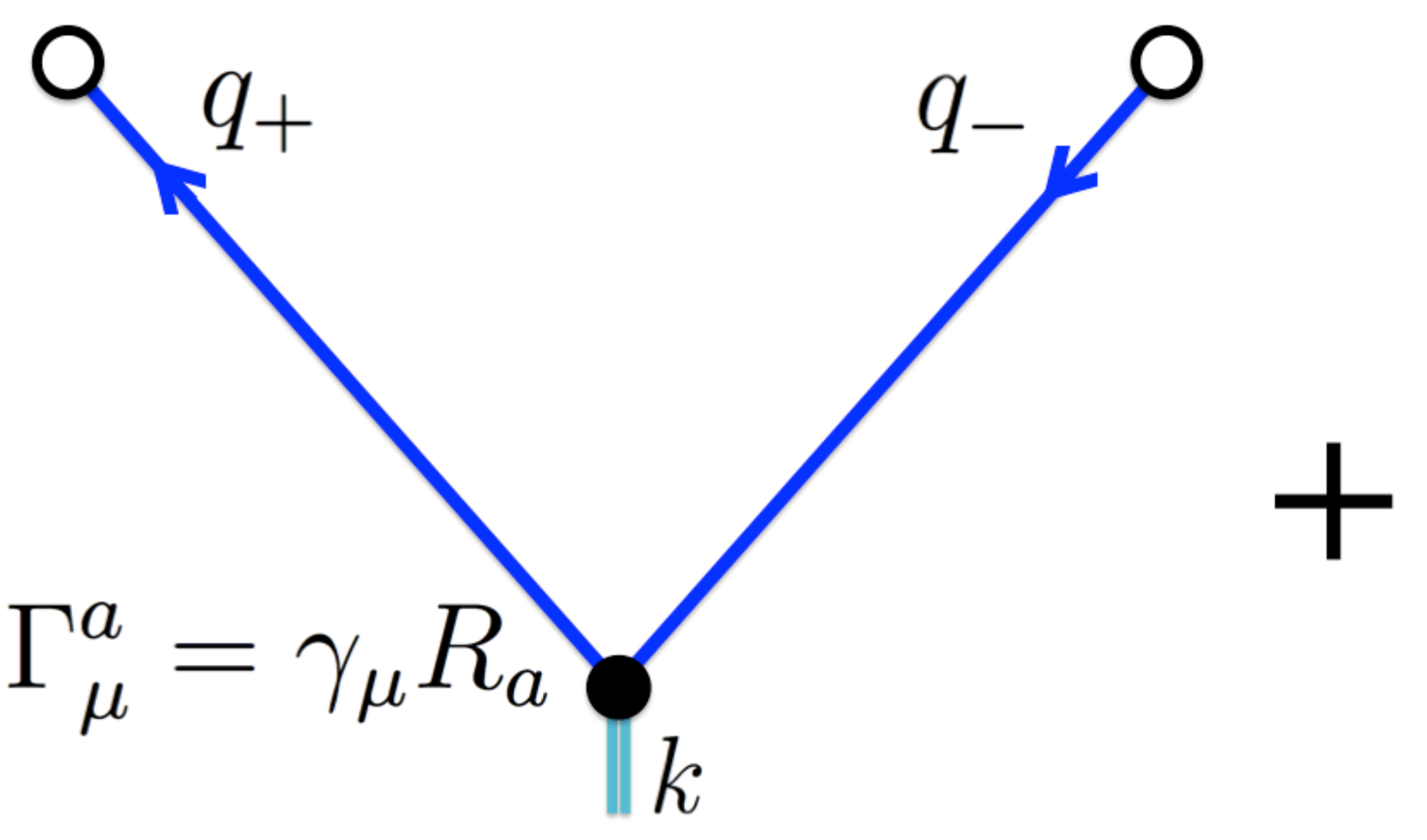} 
}\hspace{.0cm}
\scalebox{0.4}[0.4] {
  \includegraphics[scale=.43]{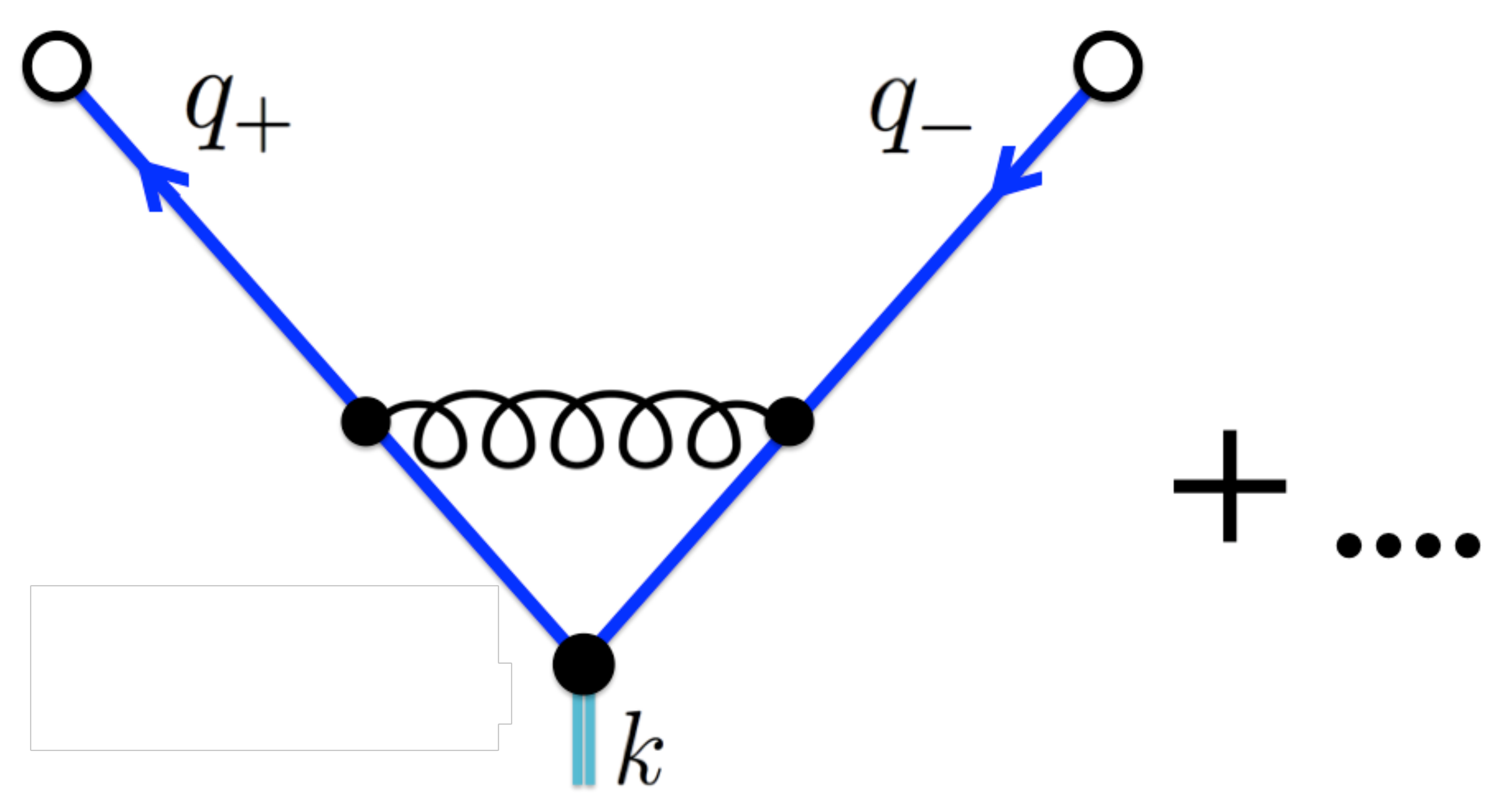} }
\end{center}
\vspace{-0.6cm}
\caption{The full vertex $\bar{ {\bf \Gamma} }_\mu^a (q_+,q_-)$
for the quark color current. The momentum fed into the vertex is 
 $k$.}
\label{fig:vertex}
\end{figure}

\begin{figure}[tb]
\vspace{0.5cm}
\begin{center}
\scalebox{0.4}[0.4] {
  \includegraphics[scale=.43]{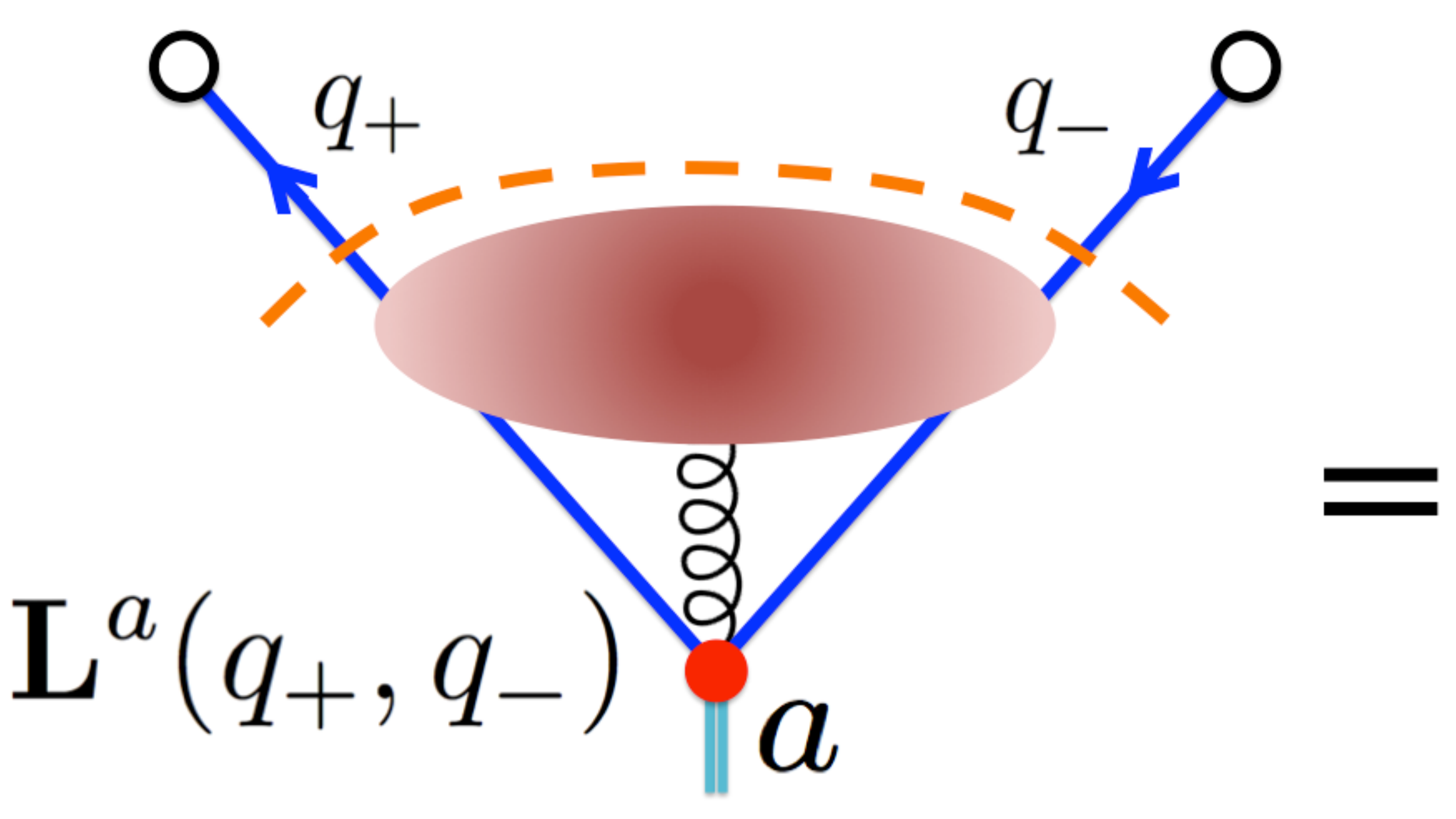} 
} \hspace{0.0cm}
\scalebox{0.4}[0.4] {
  \includegraphics[scale=.43]{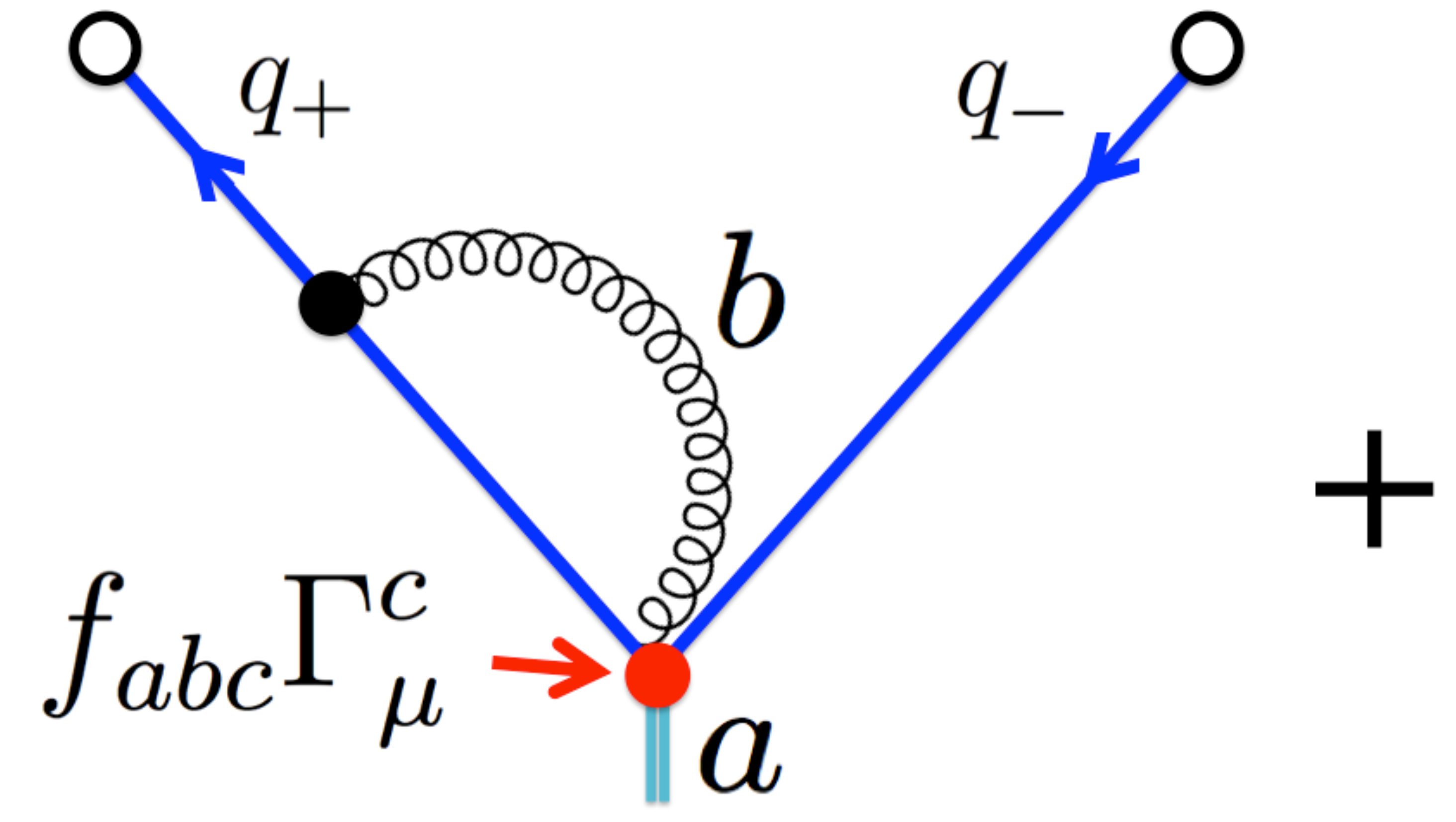} 
}\hspace{.0cm}
\scalebox{0.4}[0.4] {
  \includegraphics[scale=.43]{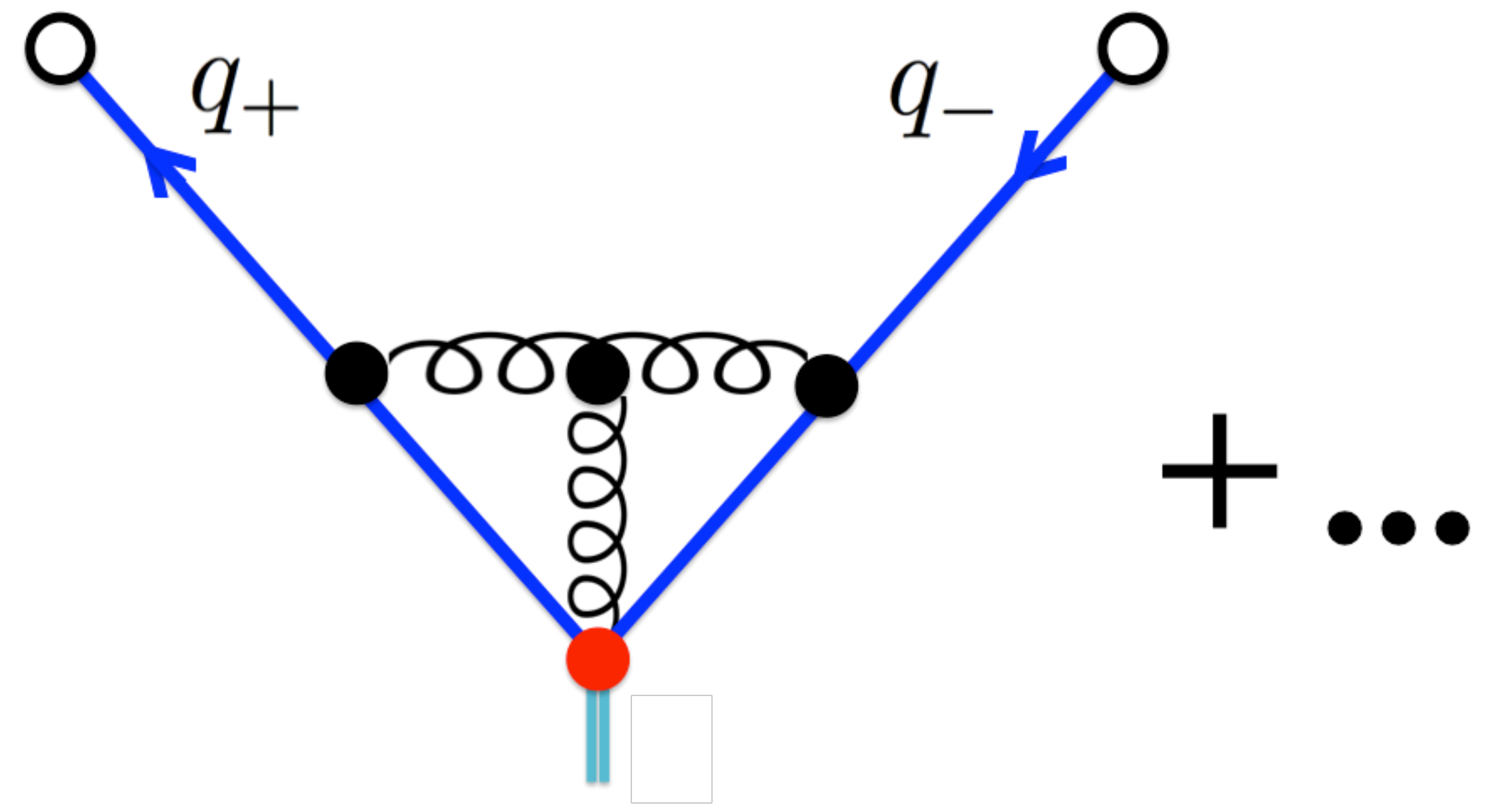} }
\end{center}
\vspace{-0.6cm}
\caption{The full vertex for the quark-gluon composite
operators $\bar{ {\bf L} }_a$.
The leading order already contains the loop structure
due to the necessity of closing the gluon lines.
All color indices except ``$a$'' are contracted.}
\label{fig:compo}
\end{figure}

In order to identify such gauge-variant contributions, 
we use a general identity
obeyed by correlation functions relating the quark color current
with other fields. The identity, whose derivation is given in the Appendix,
is
%\begin{widetext}
\beq
\left\la  D^{ac}_\mu j^c_\mu (x)  
\Psi(z_1) \bar{\Psi}(z_2) \right\ra
= - \delta^D (x-z_1) \left\la  R_a \Psi(z_1) \bar{\Psi}(z_2) \right\ra  \nonumber\\
+ \delta^D (x-z_2)  \left\la \Psi(z_1) \bar{\Psi}(z_2) R_a \right\ra.  \nonumber\\
\label{WTI}
\eeq
%\end{widetext}
%
Assuming translational invariance, we write
\beq  
\left\la\,  j^a_\mu (x) \Psi(z_1) \bar{\Psi}(z_2) \,\right\ra \nonumber\\
\equiv \int_{w,u} \calS(z_1-w)\, \bar{ {\bf \Gamma} }^a_\mu(w-x, x-u) \, 
\calS (u-z_2) \,,
\eeq
where ${\bar {\bf \Gamma} }_\mu^a$ is the full vertex
for the quark color current (Fig.~\ref{fig:vertex}), and 
\beq
\left\la\, \left( f_{abc} A_\mu^b j_\mu^c (x) \right)\, 
\Psi(z_1) \bar{\Psi}(z_2) \right\ra \nonumber\\
\equiv \int_{w,u} \calS(z_1-w)\, {\bf L}^a(w-x, x-u) \, \calS (u-z_2) \,,
\label{identity}
\eeq
where $ {\bf L}^a$ the vertex for quark-gluon composite operators
(Fig.~\ref{fig:compo}).
Note that in contrast to $\bar{ {\bf \Gamma} }_\mu^a$,
this vertex contains one loop 
already at leading order,
because the gluon line must be attached to
one of quark lines.
Taking the Fourier transform
($q_\pm = q\pm k/2$), we derive the identity which we use in the following:
\beq
\rmi k_\mu \calS(q_+) \bar{ {\bf \Gamma} }_\mu^a(q_+,q_-) \calS(q_-) \nonumber\\
= R_a \calS(q_-) - \calS(q_+) R_a 
+ \calS(q_+) {\bf L}^a(q_+,q_-) \calS(q_-) \,.
\label{identity1}
\eeq
First, we multiply Eq. (\ref{identity1}) on the left 
by $\calS^{-1} (q_+)$ and on the right by $\calS^{-1} (q_-)$
to find
\begin{align}
\rmi k_\mu \bar{ {\bf \Gamma} }_\mu^a(q_+,q_-) 
= \calS^{-1} (q_+) R_a - R_a \calS^{-1} (q_-)
+ {\bf L}^a (q_+,q_-) \,. 
\end{align}
which is the constraint for the longitudinal part
of the full vertex for quark color currents.

Next we multiply Eq. (\ref{identity1}) by $\Gamma_\nu^b = \gamma_\nu R_b$,
take the trace over color, flavor, Dirac, and Nambu-Gor'kov space,
and integrate over the momentum $q$.
The first term yields
the full current-current correlator 
$\bar{\Pi}_{\mu \nu}^{ab} (k)$,
\beq
\bar{\Pi}_{\mu \nu}^{ab} (k)
\equiv \int_q \tr \left[ \calS(q_+) \bar{ {\bf \Gamma} }_\mu^a(q_+,q_-) 
\calS(q_-) \Gamma_\nu^b \right] \,,
\label{fullcorrelator}
\eeq
and we find
%\begin{widetext}
%
\begin{align}
&\hspace{-10pt} 
\rmi k_\mu \bar{\Pi}_{\mu \nu}^{ab}(k)& \nonumber\\
&= \int_q \tr \left[\, 
\left(\, \calS(q_-) R_bR_a- \calS(q_+) R_a R_b\,\right) \gamma_\nu \, \right] \nonumber\\
&+ \int_q \tr\left[\, 
\calS(q_+) {\bf L}^a(q_+,q_-) \calS(q_-) \Gamma^b_\nu \,\right]
\nonumber \\
&=
\frac{\, \Nf \delta_{ab} \,}{2} \int_q \tr_{D,G} \left[\, 
\left(\, \calS^D (q_-) - \calS^D (q_+) \,\right) \gamma_\nu \, \right] \nonumber\\
&+ \int_q \tr\left[\, 
\calS(q_+) {\bf L}^a(q_+,q_-) \calS(q_-) \Gamma^b_\nu \,\right]\,.
\label{weget}
\end{align}
%\end{widetext}
%
which is the constraint for the longitudinal part of
the current-current correlator.

The above equations contain the vertices
specific to non-Abelian theories, ${\bf L}^a$,
which are not present in the Abelian case.
Except in the one-loop polarization function,
${\bf L}_a$ does not contribute, because
its leading order already has a one-loop structure.
This term must be included
in two-loop polarization functions.
With a  gauge-invariant regularization we find relations
for the quark current-current correlator:
\begin{align}
&k_\mu \bar{\Pi}_{\mu \nu}^{ab}(k) \big|_{1-{\rm loop} } =0\,,
\nonumber \\
&k_\mu \bar{\Pi}_{\mu \nu}^{ab}(k) \big|_{n-{\rm loop} } 
\nonumber \\ 
&= \int_q \tr\left[\, 
\calS(q_+) {\bf L}^a(q_+,q_-) \calS(q_-) \Gamma^b_\nu \,\right]
\big|_{n-{\rm loop} } ~~~ (n\ge 2).
\label{loops}
\end{align}
Beyond one loop, a number of interference terms 
among quarks, gluons, and ghosts
appear;  the sum of these terms satisfies
the transversality condition.

From this observation, we conclude that
the gauge-variant contributions which we encounter
in the simplest one-loop calculations 
must be eliminated by the gauge-invariant regularization and inclusion of the proper vertices,
{\it not} by the non-Abelian contributions related to ${\bf L}_a$.

%%%%%%%%%%%%%%%%%%%%%%%%%%%%%%%%%%%%%%
\subsection{Minimal improvement of the vertices}
%%%%%%%%%%%%%%%%%%%%%%%%%%%%%%%%%%%%%%

Let us look at the structure
of the Abelian analog of the vertex, defined by  
\beq
\rmi k_\mu{\bf \Gamma}_{\mu a}^{\calA} (q_+,q_-) 
\equiv \calS^{-1} (q_+) R_a - R_a \calS^{-1} (q_-) \,.
\eeq
Inserting the explicit expression (\ref{S}) for $\calS$,
we have
%
%\begin{widetext}
\beq
&\rmi k_\mu{\bf \Gamma}_{\mu a}^{\calA} (q_+,q_-) \nonumber\\
&=
\left[
\begin{matrix}
~ -\rmi \Slash{k} T_a~ &
\delta \Delta_c(k_+,k_-)  \gamma_5 \tau_2 \sigma_2 T_a^T ~\\
~ \gamma_5 \tau_2 \sigma_2 T_a \delta \Delta (k_+,k_-) ~
 & \rmi \Slash{k} T_a^T ~
\end{matrix}
\right] \,, \nonumber\\
\label{eqver}
\eeq
%\end{widetext}
%
where %
\beq
\delta \Delta (k_+,k_-) &\equiv& \Delta (q_+)  - \Delta (q_-) \,, \nonumber\\
\delta \Delta_C (k_+,k_-) &\equiv& \Delta_C (q_+)  - \Delta_C (q_-) \,. 
\eeq
In contrast to a U(1)$_{{\rm em}}$ superconductor,
the anomalous part is given by $\Delta(q_+)-\Delta(q_-)$
instead of $\Delta(q_+)+\Delta(q_-)$.
Accordingly, the $k_\mu \rightarrow 0$ limit gives
\beq
\rmi k_\mu{\bf \Gamma}_{\mu a}^{\calA} (q_+,q_-) 
~ \rightarrow ~ 0 \,,
~~~~~~(k_\mu \rightarrow 0)
\eeq
implying that there is no massless pole in the vertex,  a reflection of the fact that
global color symmetry is not broken.  By contrast,
the vertex for a U(1)$_{ {\rm em}}$ superconductor
acquires the anomalous contribution at small $k$,
%
%\begin{widetext}
\beq
k_\mu \delta {\bf \Gamma}_{\mu a}^{\calA ( {\rm Higgs}) } (q_+,q_-) 
\sim 2 \left[
\begin{matrix}
 0 &
 \Delta_c(q)  \gamma_5~\\
~ \gamma_5 \Delta (q) 
 & 0 ~
\end{matrix}
\right] 
\eeq
so that
\beq
&&\delta {\bf \Gamma}_{\mu a}^{\calA ({\rm Higgs})} (q_+,q_-)  \nonumber\\
&&\sim 2 \,\frac{\, k_0 g_{\mu 0} + v^2 k_j g_{\mu j} \,}
{\, k_0^2 + v^2 \vk^2 \,} \left[
\begin{matrix}
~ 0 & \Delta_c(q) \gamma_5~\\
~ \gamma_5 \Delta (q)  ~ & 0 ~
\end{matrix}
\right] \,,
\label{footnote}
\eeq
%\end{widetext}
%
where $v$ is the velocity of the massless modes in the medium.

For quark propagators with a constant gap, the constraint is satisfied with 
the bare vertex, so one can set
\beq
{\bf \Gamma}_{\mu a}^{\calA} (q_+,q_-) 
= \Gamma_\mu^a =\gamma_\mu R_a 
~~~~~~({\rm constant~gaps}).
\label{forconstant}
\eeq
This is a reflection of the fact that 
the momentum-independent self-energy
is invariant under a local color transformation.
Therefore in this case, the use of the bare vertex is not the source
of the gauge-variant contributions 
in the one-loop polarization functions.

For the momentum-dependent gaps,
the structure of the improved vertex
is much more complicated,
and so here we consider only $k_\mu \sim 0$ limit.
Expanding the left and right sides of  Eq.~ (\ref{eqver}) for small $k$
and equating terns, we have
\beq
{\bf \Gamma}_{\mu a}^{\calA} (q,q) 
=
\left[
\begin{matrix}
~ \gamma_\mu T_a~ &
\frac{\, \partial \Delta_C (q) \, }{\partial q_\mu}  
\gamma_5 \tau_2 \sigma_2 T_a^T ~\\
~ \gamma_5 \tau_2 \sigma_2 T_a 
\frac{\, \partial \Delta (q) \, }{\partial q_\mu}   ~
 & - \gamma_\mu T_a^T ~
\end{matrix}
\right]  \nonumber\\
\equiv \Gamma_\mu^a + \delta \Gamma_\mu^a (q)
~~~({\rm for~}q{\rm-dep.~gaps}).
\label{momdepvertex}
\eeq
While the diagonal component contains the bare vertex,
the anomalous part contains nontrivial
contributions proportional to the momentum
derivative of the gap function.
Note that because we are assuming that the
gap functions depend only on spatial momenta,
we have $\delta \Gamma^a_0 (q) = 0$
for the $\mu=0$ component.

As mentioned earlier,
the gap functions damp in the UV region
so that the counterterms in vacuum are
the same as for $\mu\neq 0$; thus,
one need not to worry about the regularization artifacts.
Instead, 
the gauge-variant contributions in the one-loop polarization 
functions arise from the use of the bare vertex,
and are eliminated with an improved vertex.

%%%%%%%%%%%%%%%%%%%%%%%%%%%%%%%%%%%%%%
\section{One-loop results with bare vertex}
\label{1loop}
%%%%%%%%%%%%%%%%%%%%%%%%%%%%%%%%%%%%%%

\begin{figure}[tb]
\vspace{0.0cm}
\begin{center}
\scalebox{0.5}[0.5] {
  \includegraphics[scale=.40]{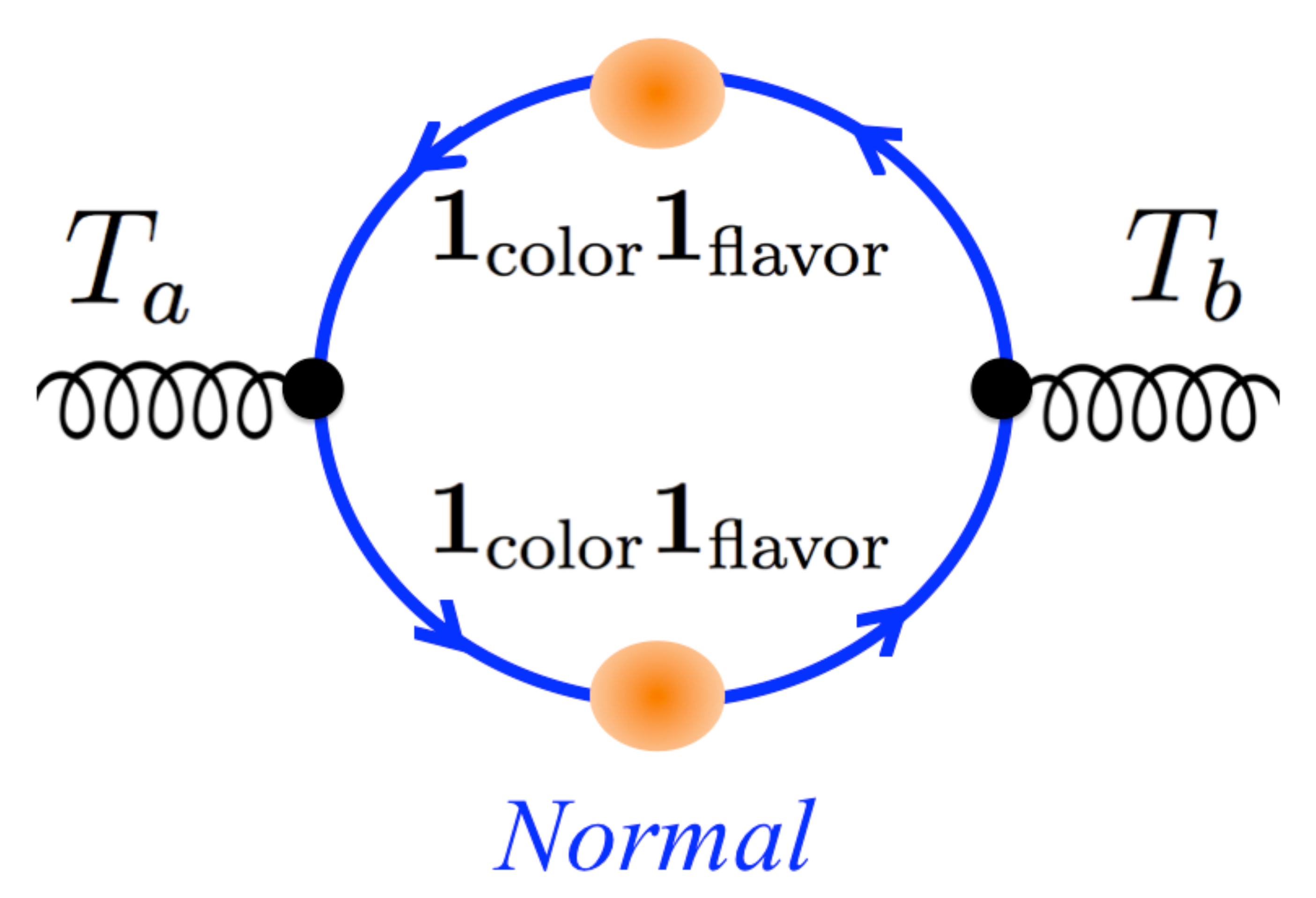} 
} \hspace{1.0cm}
\scalebox{0.5}[0.5] {
  \includegraphics[scale=.40]{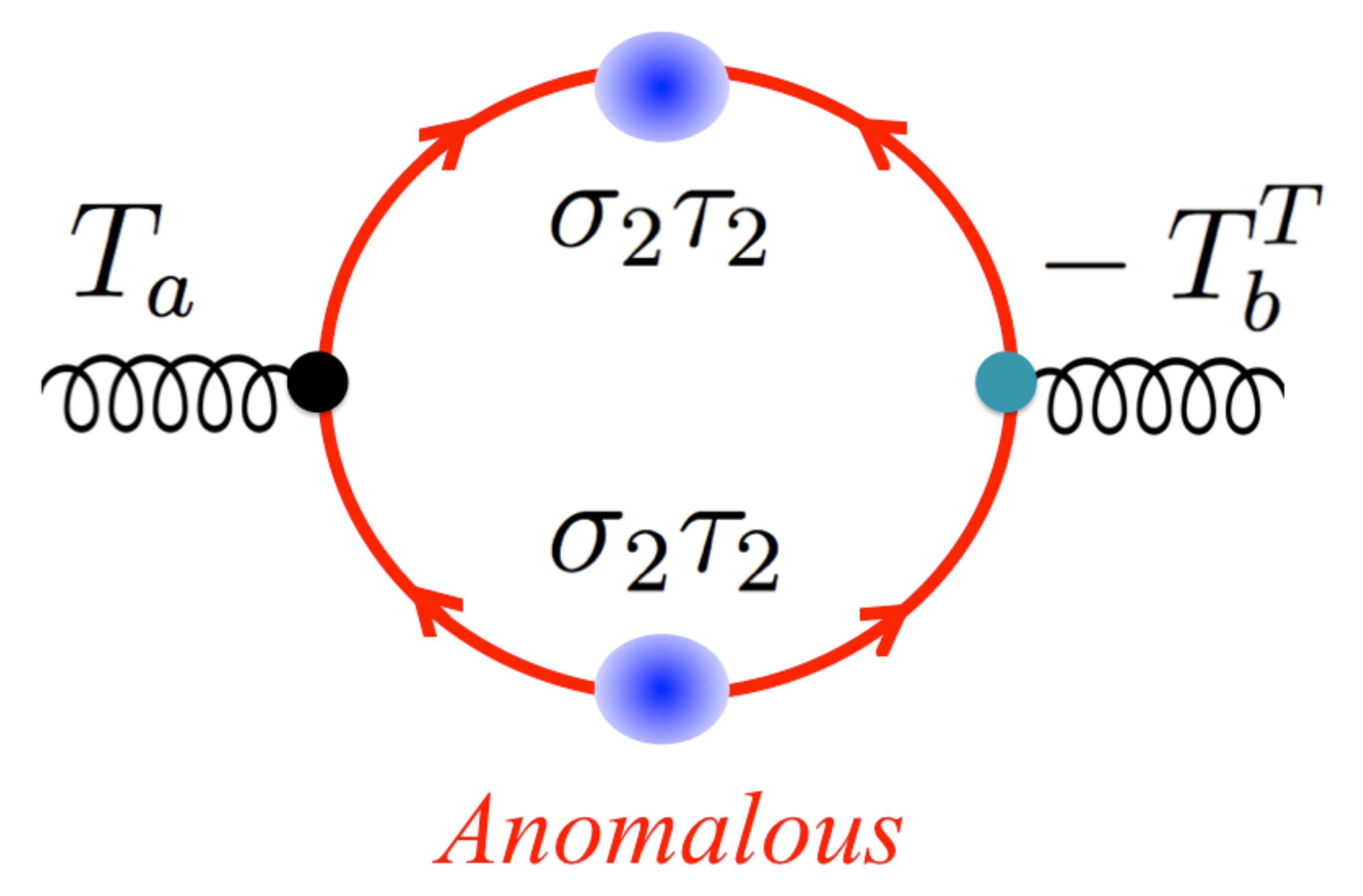} 
}
\end{center}
\vspace{-0.6cm}
\caption{The one-loop diagrams 
(where $T_a = \sigma_a/2$ for colors and
$\tau_f$ for flavors). 
Only the color-flavor structures are made explicit:
the normal loop diagrams are a product of the normal components,
$\calS_{11}\calS_{11}$ or $\calS_{22}\calS_{22}$.
The combination of the
vertices are either $(T_a, T_b)$ or $(-T_a^T, - T_b^T)$, with
only the first one shown.   The anomalous loop diagrams are a product of the anomalous components,
$\calS_{12}\calS_{21}$ or $\calS_{21}\calS_{12}$.
The combination of the
vertices is either $(T_a, -T^T_b)$ or $(-T_a^T, T_b)$.
For the U(1)$_{ {\rm em} }$ Higgs phase,
we set $T_a, \tau_2 \sigma_2 \rightarrow 1$
and the external lines are photons.
The difference from a U(1)$_{ {\rm em}}$ superconductor 
arises through the matrix elements of 
the anomalous part;
the signs of the anomalous contributions are opposite.}
\label{fig:loop}
\end{figure}

In this section we calculate the one-loop polarization function calculated
with the bare vertex and with 
a spatial momenta cutoff at $\Lambda$.
Later, in Sec.\ref{improvement},
we incorporate corrections to recover 
gauge invariance.  Explicitly (see Fig.\ref{fig:loop})
\begin{widetext}
\begin{align}
\Pi_{\mu \nu}^{ ab } (k)
&= - \int_{q} 
\tr_{c,f,D,G} \left[ \Gamma^a_\mu \calS (q_-) \Gamma^b_\nu \calS (q_+) \right] 
\nonumber \\
& =
- \tr_{c,f,D,G} 
\left[\begin{matrix}
~ \gamma_\mu T_a  & ~0 ~ \\
~0~ &  - \gamma_\mu T^T_a ~
\end{matrix}
\right]
\left[\begin{matrix}
~ \calS_{11} ~ & ~ \calS_{12} ~ \\
~ \calS_{21} ~ & ~ \calS_{22} ~
\end{matrix}
\right]
\left[\begin{matrix}
~ \gamma_\nu T_b  & ~0 ~ \\
~0~ &  - \gamma_\nu T^T_b ~
\end{matrix}
\right]
\left[\begin{matrix}
~ \calS'_{11}  ~& ~\calS'_{12} ~ \\
~\calS'_{21} ~ & ~ \calS'_{22} ~
\end{matrix}
\right] 
\nonumber \\
&
= -\delta_{ab}  \, \frac{\, \Nf \,}{2} \, 
\tr_{D} \!\left[\,
 \gamma_\mu \calS^D_{11} \gamma_\nu \calS'^D_{11} 
+ \gamma_\mu \calS^D_{22} \gamma_\nu \calS'^D_{22} 
+ \gamma_\mu \calS^D_{12} \gamma_\nu \calS'^D_{21} 
+ \gamma_\mu \calS^D_{21} \gamma_\nu \calS'^D_{12} 
\, \right] \,,
\label{trace}
\end{align}
\end{widetext}
where $q_\pm = q \pm k/2$, and 
we use 
$\tr_c (T_a T_b) = \tr_c (T_a^T T^T_b) = \delta_{ab}/2$
and $\tr_c (T_a \sigma_2 T^T_b \sigma_2 ) 
= \tr_c (T_a^T \sigma_2 T_b \sigma_2 ) = -\delta_{ab}/2$.
Note that the signs in front of the anomalous
components are opposite those for the U(1)$_{ {\rm em} }$ case,
as is easily seen by setting 
$T_a \rightarrow 1$ and $\sigma_2 \rightarrow 1$.
This sign change introduces the significant difference between the Higgs (BCS-paired) and singlet phases
because 
in the electric sector
the normal and anomalous contributions
tend to cancel in the 
SU(2) color phase and add in the Higgs phase, 
while in the magnetic sector, they tend to add in the SU(2) phase and cancel in the 
Higgs phase.

To proceed further, we factor out the $\gamma$-matrix structure
here [see Eqs. (\ref{normalpropagator1})-(\ref{anomalouspropagator2})].
For the normal part,
$\calS^D_{11} = \calS_{11}^\rmd \Lambda_\rmd \gamma_0 
+ \calS_{11}^\rma \Lambda_\rma \gamma_0 $
and
$\calS^D_{22} = \calS_{22}^\rmd \Lambda^C_\rmd \gamma_0 
+ \calS_{22}^\rma \Lambda^C_\rma \gamma_0$,
and we have
\begin{align}
&\tr_{D} \!\left[\,
 \gamma_\mu \calS^D_{11} (q_-) \gamma_\nu \calS^D_{11} (q_+) \,\right] = \nonumber\\
& \sum_{s,s' =\rmd, \rma}
\calS_{11}^s (q_-) \calS_{11}^{s'} (q_+) 
\ \tr_D \left[\, \gamma_\mu \Lambda_s (q_-) \gamma_0 
\gamma_\nu \Lambda_{s'} (q_+) \gamma_0 \, \right],
\nonumber \\
&\tr_{D} \!\left[\,
 \gamma_\mu \calS^D_{22} (q_-) \gamma_\nu \calS^D_{22} (q_+) \,\right]= \nonumber\\
&\sum_{s,s' =\rmd, \rma}
\calS_{22}^s (q_-) \calS_{22}^{s'} (q_+) 
 \tr_D \left[\, \gamma_\mu \Lambda^C_s (q_-) \gamma_0
\gamma_\nu \Lambda^C_{s'} (q_+) \gamma_0 \right]
\,.
\end{align}
while in the anomalous part, we use
$\calS^D_{12} = \calS_{12}^\rmd \Lambda_\rmd \gamma_5 
+ \calS_{12}^\rma \Lambda_\rma \gamma_5 $
and
$\calS^D_{21} = \calS_{21}^\rmd \Lambda^C_\rmd \gamma_5 
+ \calS_{21}^\rma \Lambda^C_\rma \gamma_5 $,
and find
\begin{align}
&\tr_{D} \!\left[\,
 \gamma_\mu \calS^D_{12} (q_-) \gamma_\nu \calS^D_{21} (q_+) \,\right]= \nonumber\\
& \sum_{s,s' =\rmd, \rma}
\calS_{12}^s (q_-) \calS_{21}^{s'} (q_+) 
 \tr_D \left[\, \gamma_\mu \Lambda_s (q_-) \gamma_5
\gamma_\nu
\Lambda_{s'}^C (q_+) \gamma_5\, \right],
\nonumber \\
&\tr_{D} \!\left[\,
 \gamma_\mu \calS^D_{21} (q_-) \gamma_\nu \calS^D_{12} (q_+) \,\right]= \nonumber\\
& \sum_{s,s' =\rmd, \rma}
\calS_{21}^s (q_-) \calS_{12}^{s'} (q_+) 
\tr_D \left[\, \gamma_\mu \Lambda^C_s (q_-) \gamma_5
\gamma_\nu \Lambda_{s'} (q_+) \gamma_5\, \right].
\end{align}
We write the kinematic factors for
the normal and anomalous parts as
\beq
N^{ss'}_{\mu \nu} 
\equiv  \tr_D \left[\, \gamma_\mu \Lambda_s (q_-) \gamma_0 
\gamma_\nu \Lambda_{s'} (q_+) \gamma_0 \, \right] \,, \nonumber \\
A^{ss'}_{\mu \nu} 
\equiv  \tr_D \left[\, \gamma_\mu \Lambda_s (q_-) \gamma_5 
\gamma_\nu \Lambda^C_{s'} (q_+) \gamma_5 \, \right] \,,
\eeq
from which all of remaining components can be obtained
by noting that 
$\Lambda_{\rmd,\rma} = \Lambda^C_{\rma,\rmd}$.

 In this way, the computations for the kinematic factors 
and for the $q_0$ integral of the propagator part
---which yields the ``coherence factor"---factorize.
As we will see, the structure of the polarization function takes
the simple form ($\Pi^{ab}_{\mu \nu} = \delta_{ab} \Pi_{\mu \nu}$)
\beq
\Pi_{E,M} (k)
= - \frac{\, \Nf \,}{2} \sum_{s,s'=\rmd,\rma}
 \int_{\vq} K_{E,M}^{ss'} (\vq_-, \vq_+) I^{ss'}_{E,M}
\label{structure}
\eeq
where
\beq
I^{ss'}_{E,M}\equiv
\int_{q_0}
\left[ \left( \calS^s_{11} \calS'^{s'}_{11} 
+ \calS^s_{22} \calS'^{s'}_{22} \right)
\mp \left( \calS^s_{12} \calS'^{s'}_{21} 
+ \calS^s_{21} \calS'^{s'}_{12} \right)
\right], \nonumber\\
\label{Iss}
\eeq
where the upper sign is for the electric and the lower for the magnetic response, and 
the 
kinematic factors $K$---which 
can be written in terms of $N_{\mu \nu}$ and $A_{\mu \nu}$---are common 
for the normal, Higgs,
and color-singlet phase.
On the other hand, the matrix elements---whose color and flavor structure 
we have already partially computed---and the coherence factors
reflect differences among three phases.
The results for the U(1)$_{{\rm em}}$ superconductor
are obtained by flipping the sign (lower sign) in front of 
the anomalous part in Eq. (\ref{structure}).

%%%%%%%%%%%%%%%%%%%%%%%%%%%%%%%%%%%%%%
\subsection{The kinematic factors}
\label{kinematic factors}
%%%%%%%%%%%%%%%%%%%%%%%%%%%%%%%%%%%%%%

For the computation of the electric sector,
we need the $\mu=\nu=0$ components
(in gauge-invariant computations).
The electric kinematic factor is [$E_\pm = E(q_\pm)$]
\beq
K_E^{ss'}
\equiv \frac{k^2}{\, \vk^2 \,} K_{00}^{ss'} ,
\eeq
[see Eq,~ (\ref{projection})] where
\beq
K^{ss'}_{00} \equiv  N^{ss'}_{00} = - A^{ss'}_{00} 
= 1 \pm \frac{\, \vq_- \cdot \vq_+ + m^2 \,}{E_- E_+ } ,
\label{00comp}
\eeq
with $+$ and $-$ corresponding to $s=s'$ and $s\neq s'$,
respectively.
In Eq. (\ref{00comp}) the normal and anomalous parts
have the same magnitude but opposite sign. 
Thus in the soft $\vk \rightarrow 0$ limit ,
$K^{\rmd \rmd=\rma \rma}_{00} \rightarrow 2$ while
$K^{\rmd \rma=\rma \rmd}_{00} \rightarrow 0$,
implying that---purely due to kinematic effects---the particle-antiparticle contributions
are negligible. This vanishing does not hold for the magnetic sector.

Similarly, when both indices are spatial we obtain
\beq
K_{ij}^{ss' } 
&\equiv &N^{ss'}_{ij} = A^{ss'}_{ij} \nonumber\\
&&= - \delta_{ij} \pm 
\frac{\,\delta_{ij} \left( \vq_- \cdot \vq_+ + m^2 \right)
- 2  q_{-i} q_{+j} \,}{\, E_-E_+},
\eeq
again with $+$($-$) corresponding to $s=s'$ ($s\neq s'$);
here the normal and anomalous parts have the same sign.
From this expression we project out the
magnetic and static longitudinal components defined by
\begin{align}
K_M^{ss'} 
\equiv \frac{1}{\, 2 \,} P_{ij}^M K_{ij}^{ss'} \,,
~~~~
K_{Ls}^{ss'} 
\equiv \frac{\, k_i k_j\,}{\vk^2} K_{ij}^{ss'} \,.
\label{magcomp}
\end{align}
The latter will be used to identify
the gauge-variant contributions hidden in
the magnetic sector,
see Sec.\ref{gaugemagmass}.

Finally we consider the kinematic factor for the vacuum part.
As outlined in Sec. \ref{sec:vacsubtraction}, we will compute
the vacuum part 
using the constitutent quark mass $M_\chi$.
Since the vacuum part is analogous to 
particle-antiparticle contributions,
we replace $m$ in the above kinematic factors 
$K^{\rmd \rma }= K^{\rma \rmd}$ by $M_\chi$,
\beq
K_{E,M}^{\rmd \rma} (m) &=& K_{E,M}^{\rma \rmd} (m) \nonumber\\
&\to& K_{E,M}^{ {\rm vac} } 
\equiv K_{E,M}^{\rmd \rma}  (M_\chi) =K_{E,M}^{\rma \rmd}  (M_\chi)\,, \nonumber\\
K_{Ls}^{\rmd \rma} (m) &=&K_{Ls}^{\rma \rmd} (m)
\to  K_{Ls}^{ {\rm vac} } 
\equiv K_{Ls}^{\rmd \rma}  (M_\chi) = K_{Ls}^{\rma \rmd}  (M_\chi)\,,\nonumber\\
\label{vackinematic}
\eeq
where we will need $K_{Ls}^{ {\rm vac} }$
because in the residue computations
the vacuum part also acquires 
gauge-variant components. 

%%%%%%%%%%%%%%%%%%%%%%%%%%%%%%%%%%%%%%
\subsection{The coherence factors}
%%%%%%%%%%%%%%%%%%%%%%%%%%%%%%%%%%%%%%

Having just verified that
the kinematic factors have common magnitudes
but different signs for the normal and anomalous parts,
we turn to the propagator part, Eq.~(\ref{Iss}), in which
the roles of the anomalous part 
are opposite that for the magnetic and electric sectors.
We separately discuss the
particle-hole, particle-antiparticle, and 
antiparticle-antihole contributions
(more precisely by ``particle'' 
we actually mean ``quasiparticle'').

%%%%%%%%%%%%%%%%%%%%%%%%%%%%%%%%%%%%%%
\subsubsection{Particle-hole contributions}
%%%%%%%%%%%%%%%%%%%%%%%%%%%%%%%%%%%%%%

We illustrate the calculations
for the particle-hole contributions,
taking $s=s'=\rmd$.
The normal component can be computed as follows.
For $\calS_{11} \calS'_{11}$, we have
\begin{align}
\int_{q_0} \calS^\rmd_{11} \calS'^\rmd_{11}  
= - \left[
\frac{\,  \left| u_\rmd (q_-) \right|^2 \left| v_\rmd (q_+)  \right|^2 \,}
{\, \rmi k_0 + \epsilon_\rmd(q_-) + \epsilon_\rmd(q_+) \,} \right.\nonumber\\ \left.
+ \frac{\,  \left| v_\rmd (q_-) \right|^2 \left| u_\rmd (q_+)  \right|^2 \,}
{\, - \rmi k_0 + \epsilon_\rmd(q_-) + \epsilon_\rmd(q_+) \,} 
\right] \,.
\end{align}
The result for $\calS_{22} \calS'_{22}$ can
be obtained by swapping $u$ and $v$.
Adding these two contributions, we find
\beq
\int_{q_0} \left[\, 
\calS^\rmd_{11} \calS'^\rmd_{11}  
+ \calS^\rmd_{22} \calS'^\rmd_{22} 
\, \right] 
= -2\, C_{N}^{\rmd \rmd} \calP_{\rmd \rmd} (q_-,q_+)\,, \nonumber\\
~~~~
\calP_{\rmd \rmd} (q_-,q_+) \equiv
\, \frac{ \epsilon_\rmd(q_-) + \epsilon_\rmd(q_+) }
{\, k_0^2 + \left[\epsilon_\rmd(q_-) + \epsilon_\rmd(q_+) \right]^2 \,} \,, 
\label{nocohe}
\eeq
where the coherence factor for the normal component is
\beq
C_{N}^{\rmd \rmd} (q_-,q_+) 
\equiv \left| u_\rmd (q_-) \right|^2 \left| v_\rmd (q_+)  \right|^2  \nonumber\\
+  \left| v_\rmd (q_-) \right|^2 \left| u_\rmd (q_+)  \right|^2. \eeq
Similarly 
\beq
\int_{q_0} \left[\, 
\calS^\rmd_{12} \calS'^\rmd_{21}  
+ \calS^\rmd_{21} \calS'^\rmd_{12} 
\, \right] 
= -2\, C_{A}^{\rmd \rmd} \calP_{\rmd \rmd}(q_-,q_+), \eeq
where the coherence factor for the anomalous component is
\beq
C_{A}^{\rmd \rmd} (q_-,q_+)  
\equiv \nonumber\\
u^*_\rmd v^*_\rmd (q_-) \, u_\rmd v_\rmd (q_+) 
+ u_\rmd v_\rmd (q_-) \, u^*_\rmd v^*_\rmd (q_+), \nonumber\\
\label{anocohe}
\eeq
With Eqs. (\ref{nocohe}) and (\ref{anocohe}),
we find the particle-hole contribution
for the electric and magnetic sectors
\beq
I_{E,M}^{\rmd \rmd} ( q_-, q_+) 
= -2 C^{\rmd \rmd}_{E,M}  \calP_{\rmd \rmd} (q_-,q_+) \,,
\label{iem1}
\eeq
with the coherence factors
\beq
C_{E,M}^{\rmd \rmd} (q_-,q_+) 
\equiv C_{N}^{\rmd \rmd} (q_-,q_+) \mp C_{A}^{\rmd \rmd} (q_-,q_+)  \nonumber\\
=
\left|\, u_\rmd (q_-) v_\rmd^* (q_+) 
\mp v^*_\rmd (q_-)  u_\rmd (q_+) \,\right|^2 \,.
\label{coheEMpp}
\eeq
The results for the U(1)$_{ {\rm em} }$ case can be obtained
by interchanging $E$ and $M$, enabling ready comparison of
the results  for the Higgs and singlet phases.

We summarize the characteristic features
of the particle-hole contributions
in the limit of soft momenta, $k_\mu \rightarrow 0$,
emphasizing the difference between the singlet phase and
the normal and Higgs phases.

(i) The coherence factor
for the electric sector vanishes.
Expanding $C_E^{\rmd \rmd}$ in $|\vk|$, 
we find the infrared behavior
\beq
C_E^{\rmd \rmd} 
\simeq
\left| u_\rmd (q) \right|^2 \left| v_\rmd (q) \right|^2
\left( \frac{\, \vk\cdot \vq \,}{ E_q \epsilon_\rmd(q) }
\right)^2  \nonumber\\
=
\vk^2 \cos^2 \theta_{q,k} \, \frac{\, \vq^2 \,}{\, E_q^2 \,}
\frac{\, \left| \Delta_\rmd(q) \right|^2 \,}
{\, 4\epsilon^4_\rmd(q) \,}
\,,
\eeq
where $\theta_{q,k}$ is the angle between $\vq$ and $\vk$.
Around $|\vq| \sim p_F$ or $E(q) \sim \mu$,
the expansion is equivalently one in powers of
$\vk^2/\Delta^2_\rmd (p_F)$.
The coherence factor is enhanced
for forward and backward scattering,
for which $|\cos\theta_{q,k} |\sim 1$.
Note that in the Higgs phase this IR
suppression, $\sim  \vk^2$, occurs
in the magnetic sector, instead of the electric sector;
the suppressed particle-hole contributions
fail to cancel the particle-antiparticle (diamagnetic) contributions,
yielding the Meissner effect.

(ii) At small $|\vk|$
the coherence factor for the magnetic sector behaves as
\beq
C_M^{\rmd \rmd} 
\simeq
4 \left| u_\rmd (q) \right|^2 \left| v_\rmd (q) \right|^2
=
\frac{\, \left| \Delta_\rmd(q) \right|^2 \,}
{\, \epsilon^2_\rmd(q) \,}
\,,
\eeq
and it remains $O(1)$ near the Fermi surface.
Thus finite  (paramagnetic) contributions will cancel
the diamagnetic contributions,
as in a normal conductor.
In the Higgs phase, the IR contribution is finite
in the electric sector and rise to a finite Debye mass.

(iii) The propagator part $\calP_{\rmd \rmd}$
has an IR cutoff near the Fermi surface, as a consequence of the gap $\Delta_\rmd$.
In contrast, in the normal phase, the vanishing behavior of $C_E^{\rmd \rmd}$
is compensated by the vanishing denominator of $\calP_{\rmd \rmd}$,
yielding a finite Debye mass.
In the singlet case, $C_E^{\rmd \rmd} \rightarrow 0$
but $\calP_{\rmd \rmd}$ stays finite,
preventing an electric mass.
While the coherence factor behaves similarly in
the normal and singlet phases,
the differing behaviors of $\calP_{\rmd \rmd}$ create
the essential difference between the two phases.

(iv) The existence of the gap also 
suppresses Landau damping,
since the allowed phase space for the decays 
is very small for small $|\vk|$.
To see this, it is useful to recall that 
in the normal or gapless phase, an
expansion of $\calP^{ {\rm normal} }_{\rmd \rmd}$ 
in $k^2_0$ is ill defined due to
singular contributions from small angles; the expansion actually starts with $k_0/|\vk|$
[recall that in our metric, Euclidean and Minkowski momenta
are related as $k_0=(k_0)_E = \rmi (k_0)_M$],
\beq
\sim 
\int_{-1}^1 \frac{ \, {\rm d} \cos \theta  \, }
{\, k_0^2 + \vk^2 \cos^2 \theta \,} 
\sim -\rmi \, \frac{\, k_0^2 \,}{|\vk|^2} 
\ln \frac{ k_0 + \rmi |\vk| }{\,  k_0 - \rmi |\vk| \,}
~\rightarrow~ \frac{\, k_0 \,}{|\vk|} \nonumber\\
(|\vk| \gg k_0;~ {\rm normal~phase}). \nonumber\\
\eeq
However in the singlet case small angle scattering is not singular
at small $|\vk|$,
so that an expansion of $\calP_{\rmd \rmd}$ in powers of
$k_0^2/|\Delta|^2$ is well defined, and does not produce terms linear in $k_0$.
Thus in the gapped phase Landau damping effects 
appear only for $|\vk| \gg \Delta_{\rmd \rmd}$.

%%%%%%%%%%%%%%%%%%%%%%%%%%%%%%%%%%%%%%
\subsubsection{Antiparticle-antihole contributions}
%%%%%%%%%%%%%%%%%%%%%%%%%%%%%%%%%%%%%%

In a diquark condensate,
the Dirac sea is not fully occupied, so 
there are the antiparticle-antihole contributions to the polarization.
These contributions can be readily obtained 
by replacing the index ``$\rmd$'' with ``$\rma$''
in the result (\ref{iem1}) for the particle-particle contributions,
\beq
I_{E,M}^{\rma \rma} ( q_-, q_+) 
= -2 C^{\rma \rma}_{E,M} \calP_{\rma \rma} (q_-,q_+), \nonumber\\
C_{E,M}^{\rma \rma} (q_-,q_+) 
\equiv
\left|\, u_\rma (q_-) v_\rma^* (q_+) 
\mp v^*_\rma (q_-)  u_\rma (q_+) \,\right|^2 \,.
\label{coheEMaa}
\eeq
In the coherence factor, 
both the first and second terms in the bracket
contain $v_a$, so the antiparticle-antiparticle
contributions are suppressed for $\mu \gg \Delta_\rma$,
and are of order $\sim \Delta_\rma^2/\mu^2$.

%%%%%%%%%%%%%%%%%%%%%%%%%%%%%%%%%%%%%%
\subsubsection{Particle-antiparticle contributions
and vacuum subtraction}
%%%%%%%%%%%%%%%%%%%%%%%%%%%%%%%%%%%%%%

The particle-antiparticle
contributions are rather insensitive to  
condensation near the Fermi surface.
On the other hand, the particle-antiparticle contributions
are UV divergent, so we carefully consider the vacuum contribution as well.
Taking $s=\rmd$ and $s'=\rma$, we have
\beq
I_{E,M}^{\rmd \rma} ( q_-, q_+) 
= -2\, C^{\rmd \rma}_{E,M}  \calP_{\rmd \rma} (q_-,q_+) \,, \nonumber\\
\calP_{\rmd \rma} (q_-,q_+) \equiv
\, \frac{ \epsilon_\rmd(q_-) + \epsilon_\rma(q_+) }
{\, k_0^2 + \left[\epsilon_\rmd(q_-) + \epsilon_\rma(q_+) \right]^2 \,} \,, 
\label{propa}
\eeq
where the coherence factor is
\beq
C_{E,M}^{\rmd \rma} (q_-,q_+) 
\equiv
\left|\, u_\rmd (q_-) u_\rma^* (q_+) 
\mp v^*_\rmd (q_-) v_\rma (q_+) \,\right|^2 \,.
\label{coheEMpa}
\eeq
Similarly $I_{E,M}^{\rma \rmd}$ can be obtained
by swapping ``$\rmd$'' and ``$\rma$'' in the expression (\ref{propa}) for
$I_{E,M}^{\rmd \rma}$.
Note that for $\mu \gg \Delta_\rma$,
the second term in the bracket 
is much smaller than the first, since
$u_a \simeq 1$ and $v_a \simeq 0$
with corrections of $\sim \Delta_\rma^2/\mu^2$; thus
$ C_{E,M}^{\rmd \rma} \simeq |u_\rmd (q_-)|^2
\sim \theta(E(q_-)-\mu)$.
The anomalous components play little role
in the particle-antiparticle contributions.

We next derive the vacuum contribution from Eq.~(\ref{coheEMpa})
doing a parallel computation.
We first note that as $\Delta \rightarrow 0$ 
and $\mu \rightarrow 0$, one has
$ \epsilon_{\rmd,\rma} (q) ~\rightarrow ~ 
E_q^{{\rm vac} } = E_q (m\rightarrow M_\chi)$.
In addition $u_{\rmd,\rma} \rightarrow 1$ and
$v_{\rmd,\rma} \rightarrow 0$,
so that the coherence factor is simply unity.
Summing the ($\rmd,\rma$) and ($\rma,\rmd$) contributions,
we find
\beq
I_{E,M}^{ {\rm vac} } (q_-,q_+)
\equiv  -4\, \calP_{ {\rm vac} } (q_-,q_+) \,,\nonumber\\
~~~~
\calP_{ {\rm vac} } (q_-,q_+) \equiv
\, \frac{ E^{ {\rm vac} }_- + E^{ {\rm vac} }_+ }
{\, k_0^2 + \left(\, E^{ {\rm vac} }_- + E^{ {\rm vac} }_+ \, \right)^2 \,} \,.
\label{vacpa}
\eeq
This contribution will be subtracted from the
particle-antiparticle contributions.

%%%%%%%%%%%%%%%%%%%%%%%%%%%%%%%%%%%%%%
\subsection{Summary of one-loop results
with the bare vertex}
%%%%%%%%%%%%%%%%%%%%%%%%%%%%%%%%%%%%%%

Combining Eqs. (\ref{trace}), (\ref{Iss}), and (\ref{structure}), 
and the expression for $I^{ss'}$, we summarize
our results for $\Pi^{ab}_{\mu \nu} = \delta_{ab} \Pi_{\mu \nu}$:
\beq
\Pi_{E,M} (k) = \Nf \sum_{s,s'=\rmd,\rma}
\int_{\vq} K_{E,M}^{ss'} (q_+, q_-) \, \nonumber\\
\times C_{E,M}^{ss'} (q_+, q_-) \, \calP_{ss'} (q_-, q_+) \,,
\eeq
where the kinematic factors $K^{ss'}_{E,M}$ are defined 
in Eqs.~(\ref{00comp}) and (\ref{magcomp}),
the coherence factors 
$C^{ss'}_{E,M}$ are given by Eqs.~(\ref{coheEMpp}),
(\ref{coheEMaa}), and (\ref{coheEMpa});
the propagator factors 
$\calP^{ss'}$ are given by Eqs.~(\ref{nocohe}) and (\ref{propa}) 
with suitable replacements of the indices ``$\rmd$''
and ``$\rma$.''
The corresponding vacuum part 
[see Eqs.~(\ref{vackinematic}) and (\ref{vacpa})] is
\beq
\Pi^{ {\rm vac} }_{E,M} (k) 
= 2 \Nf 
\int_{\vq} K_{E,M}^{ {\rm vac} } (q_+, q_-) \,
\calP_{ {\rm vac} } (q_-, q_+) \,,
\eeq
and the finite polarization functions are
\beq
\itDelta \Pi_{E,M} (k) 
= \Pi_{E,M} (k) - \Pi^{ {\rm vac} }_{E,M} (k) \,.
\eeq
As in Eq.~(\ref{reno}), this term should be added to the renormalized vacuum polarization function
$\Pi^{R, {\rm vac} }_{E,M}$ to derive
the renormalized medium polarization function, 
$\Pi^{R }_{E,M} = \Pi^{R, {\rm vac} }_{E,M} + \itDelta \Pi_{E,M}$.

The static component of the longitudinal part takes the form
\begin{align}
&\Pi_{Ls} (k)\equiv \frac{\, k_i k_j \,}{\vk^2}\, \Pi_{ij} 
 \nonumber\\ &= \Nf \sum_{s,s'=\rmd,\rma}
\int_{\vq} K_{Ls}^{ss'} (q_+, q_-) \,
C_{M}^{ss'} (q_+, q_-) \, \calP_{ss'} (q_-, q_+) \,, \nonumber\\
&\Pi^{ {\rm vac} }_{Ls} (k) \equiv \frac{\, k_i k_j \,}{\vk^2}\Pi^{ {\rm vac} }_{ij} 
= 2\Nf 
\int_{\vq} K_{Ls}^{ {\rm vac} } (q_+, q_-) \,
\calP_{ {\rm vac} } (q_-, q_+) \,, \nonumber\\
\label{PiLs}
\end{align}
and contains the same coherence factor as the magnetic case
because $\Pi_{Ls} \propto k_i k_j \Pi_{ij}$;
the only difference comes from the kinematic factor, Eq.~(\ref{magcomp}).
In next section, we use
this fact to derive important
infrared relations between 
the magnetic and longitudinal components.

%%%%%%%%%%%%%%%%%%%%%%%%%%%%%%%%%%%%%%
\section{Corrections to one-loop results --
Recovery of gauge invariance}
\label{improvement}
%%%%%%%%%%%%%%%%%%%%%%%%%%%%%%%%%%%%%%

In this section we recover gauge invariance,
which was violated in the last section 
either by the use of the bare vertex or
gauge-variant regularization.
First we review how the magnetic and
longitudinal polarization tensors are related in the infrared limit,
and how differences among the normal, Higgs, and singlet phases arise.
Then we argue how the gauge invariance
requires the magnetic mass 
in the singlet and normal phases to be zero,
and why the Higgs phase escapes such a requirement.
We then give more a concrete discussion about
how to identify the gauge-variant part as 
an artifact of regularization.
As we shall see, the corresponding counterterm to carry out the regularization can be gap dependent 
if the gaps do not damp sufficiently fast in the UV region.

%%%%%%%%%%%%%%%%%%%%%%%%%%%%%%%%%%%%%%
\subsection{Gauge invariance and magnetic 
mass} \label{gaugemagmass}
%%%%%%%%%%%%%%%%%%%%%%%%%%%%%%%%%%%%%%

In the magnetic sector,
the particle-hole and particle-antiparticle contributions 
are comparable, and tend to cancel each other.
But they are qualitatively different contributions,
so at first sight their relation is not very clear.  Establishing their relation
is particularly important
in order to check whether or not a magnetic mass exists
in the singlet phase.

Actually, the balance between the Fermi-surface
contributions and particle-antiparticle contributions
are tightly constrained by gauge invariance.
To see this, we derive a useful relation [Eq.~(\ref{theorem1}] 
between $\Pi_M$ and $\Pi_{L}$ in the infrared. 
Below we consider the static limit ($k_0=0$)
for which $\Pi_L \rightarrow \Pi_{Ls}$
[Eq.~(\ref{PiLs})].

The relation relies on the fact that
the product of $C_{M}^{ss'}$ and $\calP_{ss'}$
does not depend on $\theta_{q,k}$
to leading order of $\vk$, a condition
satisfied in the normal, Higgs, and singlet phases.
Then at small $k$ 
the integral over the angle in the $\vq$ integration 
can be factorized,
\beq
\Pi_{M, Ls} (k\rightarrow 0) = 
\frac{\Nf}{\, 2\pi^2 \,} \sum_{s,s'=\rmd,\rma}
\int_0^\infty \!\! {\rm d} |\vq| \, |\vq|^2 \,
C_{M}^{ss'}\nonumber\\ \times \calP_{ss'} (q, q) \,
\int_0^1 {\rm d} \cos\theta \, K_{M, Ls}^{ss'} 
\left(q; \cos \theta \right) \,.
\eeq
Remarkably, when $\vk=0$,
explicit calculations for the angular integral give
\beq
\int_0^1 {\rm d} \cos\theta \, K_{M}^{ss'} \!\left(q; \cos\theta \right)
= \int_0^1 {\rm d} \cos\theta \, K_{Ls}^{ss'} \!\left(q; \cos\theta \right) \nonumber\\
({\rm for}~\vk=0)
\eeq
for any combination of $(s,s')$; 
the difference of the integrals starts with $O(\vk^2)$ contributions.
Since the coherence factor and propagator 
are common for the magnetic and longitudinal sectors,
we conclude that
\beq
\Pi_M (k\rightarrow 0) = \Pi_{Ls} (k\rightarrow 0) \,.\nonumber\\
~~~~~~(\Pi_{M,Ls}:{\rm bare~vertex~results})
\label{theorem}
\eeq
The above argument works equally well for the 
vacuum part, and
so we arrive at the same conclusion for 
$\itDelta \Pi_{M,Ls}$.
The relation holds for normal, Higgs, and singlet phases.

A nonvanishing $\Pi_{Ls}$ is purely a consequence of the
computation being gauge variant. 
First we consider how the improved vertex
reduces the problem, and will see the differing role
of the improved vertex for
phases with and without 
symmetry breaking.

In the singlet and normal phases, color symmetry is not broken; thus
the improved vertex $\delta \Gamma_\mu$
does not contain massless modes
but rather behaves as $\delta \Gamma_\mu(q,q) \propto g_{\mu j} q_j$;
see Eq. (\ref{momdepvertex}).
Then corrections from the improved vertex
$\delta_{ {\rm v} }\Pi_{\mu \nu}(k)$ are of the form $ \sim g_{\mu i} g_{\nu j}\delta_{ij}
V(k^2)$, where $V(k^2)$ is a regular function of $k^2$.
After projecting the correction onto the magnetic and longitudinal sectors,
we can see that contributions to 
the magnetic and longitudinal components are equal, and thus
\beq
\bPi_M 
= \Pi_M + V
= \Pi_{Ls} + V
= \bPi_{Ls} \,, \nonumber\\
~~~~~~(k\rightarrow 0: {\rm singlet, normal~phases}),
\label{theorem1}
\eeq
and the relation $\Pi_M(k\rightarrow 0) = \Pi_{Ls} (k\rightarrow 0)$
can be carried over to $\bPi_M(k\rightarrow 0) = \bPi_{Ls} (k\rightarrow 0)$.
If $\bar{\Pi}_{Ls}$ is still nonvanishing, 
it must be an artifact of the gauge-variant regularization, which we must
eliminate by counterterms.
As we will see later, 
the counterterm 
$\delta_{ {\rm c} } \Pi_{\mu \nu}$ again must have a tensor structure 
$\delta_{ {\rm c} }\Pi_{\mu \nu} \sim g_{\mu i} g_{\nu j}\delta_{ij}
C(k^2)$, so an attempt to erase the longitudinal component by 
a counterterm {\it precisely} eliminates a magnetic mass,
i.e.,
\beq
\Pi_M^{ {\rm phys} } 
= \bPi_M + C
= \bPi_{Ls} + C
= \Pi_{Ls}^{ {\rm phys} } = 0\,, \nonumber\\
~~~~~~(k\rightarrow 0: ~{\rm singlet,normal~phases}).
\eeq
We thus conclude that there should be no magnetic mass
in either the singlet or normal phases in a gauge-invariant computation
because of the lack of color-symmetry breaking.

In the Higgs phase, the improved vertex at $k_0=0$ adds contributions 
$\delta_{ {\rm v} }\Pi_{\mu \nu} \sim V g_{\mu i} g_{\nu j} k_i k_j/\vk\,^2$, 
reflecting the existence of the massless modes
in the vertex.
In contrast to the singlet case,
the improved vertex does not affect the magnetic sector
because the projection operator $P_{\mu \nu}^M$ 
eliminates this term.
At this stage the magnetic and longitudinal 
components are no longer equal,
\beq
\bPi_M = \Pi_M 
~\neq~ \Pi_{Ls} + V
= \bPi_{Ls} \,.\nonumber\\
~~~~~(k\rightarrow 0: ~{\rm Higgs~phases})
\label{theorem2}
\eeq
from which we conclude that after adding counterterms,
\beq
\itDelta \Pi_M^{ {\rm phys} } 
~\neq~ \itDelta \Pi_{Ls}^{ {\rm phys} } = 0\,,\nonumber\\
~~~~~(k\rightarrow 0: ~{\rm Higgs~phase})
\eeq
We see that the existence of massless modes 
totally changes the situation, allowing a magnetic mass in the Higgs phase.

%%%%%%%%%%%%%%%%%%%%%%%%%%%%%%%%%%%%%%
\subsection{Identification of regularization artifacts}
\label{subSec:constantgap}
%%%%%%%%%%%%%%%%%%%%%%%%%%%%%%%%%%%%%%

We now show how regularization via counterterms 
can violate gauge invariance, even after
an improved vertex is used.  Our primary aim is to illustrate how to
identify  counterterms and their structure.
We start from Eq. (\ref{weget})
at the one-loop level,
\beq
&\rmi k_\mu \bPi^L_{\mu \nu} (k) \big|_{1-{\rm loop} } \nonumber\\
& = \frac{\, \Nf \,}{2} \int_q \tr_{D,G} \left[\, 
\left(\, \calS^D (q_-) - \calS^D (q_+) \,\right) \gamma_\nu \, \right]
\,,
\label{tensornu}
\eeq
where $\bar{\Pi}_{\mu \nu}$ includes the 
improved vertex, (\ref{fullcorrelator}).
We first note that the right side of (\ref{tensornu}) is independent of
$k_0^2$, since in the absence of a cutoff in $q_0$
we can freely shift $q_0$ to eliminate any $k_0$ dependence;
therefore, the right side depends only on $\vk$.

For $\nu=0$, $k_\mu \bPi^L_{\mu 0} (k)$ vanishes for all $k$,
a consequence of the fact that after we take residues
the contributions from $\psi$ and $\psi_C$ 
precisely cancel for each spatial momentum\footnote{To 
avoid confusion, we emphasize that
$\tr_{D,G} \left[\calS^D \gamma_\nu \right]
= \tr_D \left[ \left(\calS_{11}^D + \calS_{22}^D \right) 
\gamma_\nu \right]$
is {\it not} the quark number current, which in
the Nambu-Gor'kov bases is instead 
$\tr_D \left[ \left(\calS_{11}^D - \calS_{22}^D \right) 
\gamma_\nu \right]$, and is non-zero for $\nu=0$.},
\beq
\int_{q_0} \tr_{D,G} \left[\, \calS^D (q) \gamma_0 \, \right]& \nonumber\\
= \int_{q_0} \tr_{D} \left[\, 
\left( \calS_{11}^D (q) + \calS_{22}^D(q) \right) \gamma_0 \, \right]
= 0 .
\eeq
Thus
\beq
k_\mu \bPi^L_{\mu 0} (k) 
= k_0 \bPi^L_{00} + k_j \bPi^L_{j0} =0 \,.
\label{rela}
\eeq
The same argument also holds for the vacuum part,
$\bPi^{ {\rm vac}, L}_{\mu 0}$.

For $\nu =j$ the terms in (\ref{tensornu})
no longer vanish.  Rather, 
\begin{align}
&\int_q \tr_{D,G} \left[\, 
\left(\, \calS^D (q_-) - \calS^D (q_+) \,\right) \gamma_j \, \right]
\nonumber \\
&= 2 \rmi \sum_{s=\rmd,\rma} 
\int_{\vq} \left[
\left(q_j -\frac{\, k_j \,}{2} \right)
\frac{\, |u_s(q_-) |^2 - |v_s(q_-) |^2 \,}{E_-} \right.\nonumber\\ &\left.
- \left(q_j + \frac{\, k_j \,}{2} \right)
\frac{\, |u_s(q_+) |^2 - |v_s(q_+) |^2 \,}{E_+} \right]
\,.
\label{su}
\end{align}
We now investigate the small-$\vk$ behavior.
Explicitly writing the rotationally symmetric UV cutoff
as $\theta(\Lambda^2-\vq^2)$,
we can rewrite the above integral as
\begin{widetext}
\begin{align}
& \int_{\vq} \left[\,  
\theta\left(\, \Lambda^2 - \left(\, \vq + \vk /2 \,\right)^2 \,\right)
- \theta\left(\, \Lambda^2 - \left(\, \vq -  \vk /2 \, \right)^2 \,\right) \,
\right] \, q_j \,
\frac{\, |u_s(q) |^2 - |v_s(q) |^2 \,}{E_q}
\nonumber \\
& = - 2 k_j \, \int_{\vq}   
\delta\left( \Lambda^2 - \vq^2 \right) \,
\frac{\, \vq^2 \,}{3} \, \frac{\, |u_s(q) |^2 - |v_s(q) |^2 \,}{E_q} + O(k^3)
\,.
\label{shiftintegral}
\end{align}
\end{widetext}
Note that at large $|\vq| \gg \mu, \Delta$,
\beq
|u_{\rmd,\rma} (q) |^2 - |v_{\rmd, \rma} (q) |^2
= \frac{E_q\mp\mu}{\sqrt{ (E_q\mp \mu)^2 + \Delta_{\rmd,\rma}^2 } } \nonumber\\
= 1 - \frac{\, |\Delta_{\rmd, \rma}|^2 \,}{2|\vq|^2 } 
+ O(1/\vq^4) \,,
\eeq
and $E_q^{-1} \simeq 1/|\vq| - m^2/2|\vq|^3$,
so that we finally identify the degree of
transversality violation:
\beq
k_\mu \bPi_{\mu j}
= k_j \left(-\, \frac{ \Nf }{\, 6 \pi^2 \,} \right) 
\sum_{s=\rmd,\rma}  
\left(\, \Lambda^2 -\frac12 ( |\Delta_s|^2 + m^2 )\right.\nonumber\\ \left.
+ O(\Lambda^{-2} )\,\right) + O(\vk\,^3) \,,
\eeq
where $\Delta_s$ is essentially
the gap function at $|\vq|=\Lambda$.
The $\Lambda^2$ term also appears 
in the vacuum contribution
and can be eliminated by the vacuum subtraction.
The second term, however, survives 
even after the vacuum subtraction and when taking the
$\Lambda \rightarrow \infty$ limit.
Thus after subtracting the vacuum contributions
with the mass gap $M_\chi$,
the gauge-variant contribution to the condensation effects is characterized by
\beq
k_\mu \itDelta \bPi_{\mu j}
= k_j \frac{ \Nf }{\, 12 \pi^2 \,}  
\sum_{s=\rmd,\rma} 
\left( \, |\Delta_s|^2 + m^2 - M_\chi^2 \, \right) 
\nonumber\\ + O(\vk^3) 
\equiv k_j C_{{\rm gaps}} (\vk^2) \,.
\label{subnormal}
\eeq
These terms, which reflect the coupling of regularization artifacts
to the gaps, must be handled individually for the different phases
whenever their gaps are not equal to those in vacuum
 \footnote{Even in normal quark matter,
this contribution should be taken into account because
the mass in the QCD vacuum, the effective mass $M_\chi$,
differs from the current mass in chirally restored normal quark matter.
Usual hard dense loop calculations 
tacitly avoid this gauge variant artifact
by using the current quark mass $m$ in the chirally symmetric vacuum.}.

Actually, in realistic treatments of gap functions in QCD,
$\Delta_s$ and $M_\chi$ damp sufficiently fast in the UV
that these problems are automatically bypassed.
Instead, it becomes necessary to improve the vertex.  

Note that the violation of the transversality condition
that we found above 
is a purely technical problem, because 
the use of a momentum cutoff did not allow
a shift in momentum.
Had we instead used  dimensional regularization 
we could have eliminated the $\Delta^2$, etc. terms 
automatically, as we can easily see from Eq.~(\ref{tensornu}).
We conclude that
the aforementioned constant terms
were introduced  purely by hand
through the regularization scheme,
and {\it must} be removed by counterterms
designed to erase the regularization artifacts.

In principle, we can imagine two types of counterterms
that could eliminate nonzero contributions 
in $k_\mu \bPi_{\mu j} \sim k_j$: the
first is proportional to $\delta_{ij}$,
and the second is proportional to 
$k_\mu k_j/k^2$.
Without color-symmetry breaking
(we postpone the discussions of the Higgs phase
to the end of this subsection),
it is easy to reject the second type of counterterm
by recalling that the Ward-Takahashi identity for the vertex function
behaves at small momenta as
\beq
k_\mu {\bf \Gamma}^a_\mu (q_+,q_-) ~\rightarrow~ 0 
\nonumber \\
~~~~~(k_\mu \rightarrow 0:~
{\rm singlet~or~normal~phase}),
\label{vertexlimit}
\eeq
implying that
the vertex does not contain any massless poles.
In fact, if there were a $k_\mu k_j/k^2$ term, 
the left side would approach a constant.
Since the only possible way to produce
masseless modes is the improved 
vertex\footnote{The 
remaining part does not contain the interaction
so that it can yield only a cut instead of poles.},
we conclude that the artificial contributions 
introduced by our regulator
do not couple to massless modes.
Therefore we do not consider $k_i k_j/k^2$-type
counterterms; and consider only counterterms proportional to $\delta_{ij}$:
\beq
\delta_{ {\rm c} } \bPi^{ {\rm gaps} }_{\mu \nu} 
= - g_{\mu i} g_{\nu j} \delta_{ij} \, C_{{\rm gaps}} (\vk^2)
\,,
\label{ijcounter}
\eeq
from which the desired transversality condition,
\beq
k_\mu \left(
 \itDelta \bPi_{\mu \nu} + \delta_{ {\rm c} } \bPi_{\mu \nu}^{ {\rm gaps} } \right) 
= k_\mu \itDelta \bPi^{ {\rm phys} }_{\mu \nu} =0\,,
\label{recovery}
\eeq
is recovered.
Multiplying Eq. (\ref{recovery}) by $k_\nu$, we have
\beq
C_{ {\rm gaps} } (\vk^2)
= \frac{\, k_\alpha k_\beta \,}{\vk^2} \itDelta \bPi_{\alpha \beta} 
= - \frac{\, k_0^2 \,}{\vk^2} \itDelta \bPi_{00} 
+ \frac{\, k_i k_j \,}{\vk^2} \itDelta \bPi_{ij}, \nonumber\\
\eeq
where we have used the relation (\ref{rela}) to eliminate
the $\bPi_{0j}$ components.
Both terms are regular in the $\vk \rightarrow 0$ limit
($\Delta \bPi_{00} \sim \vk^2$ at small $\vk$).
In this way, 
the term $C_{ {\rm gaps} }$ is uniquely 
determined \footnote{If we wish to find
the vertex correction, $\delta_{ {\rm v}} \Pi_{\mu \nu}$, one can,
instead of calculating it explicitly,
compute $C_{ {\rm gaps} }$ and
$\Pi_{\mu \nu}$ for the bare vertex, and then
use them to read off $\delta_{ {\rm v}} \Pi_{\mu \nu}$. 
In particular, when we consider the damping of gap functions in the UV,
we can set $C_{ {\rm gaps} }=0$
and directly relate $\Pi^L_{\mu \nu}$ to
$-\delta_{ {\rm v}} \Pi_{\mu \nu}$
because $\Pi^L_{\mu\nu} + \delta_{ {\rm v}} \Pi_{\mu \nu}=0$.
}.
Therefore although we introduce counterterms
that are dependent on phases, 
they produce well-defined results.

Although we introduce a counterterm to 
eliminate the gauge-variant longitudinal components,
the counterterm enters 
the results for both the electric and magnetic sectors.
The reason is that in naive computations 
the projection operators pick up
physical as well as artificial contributions
having a tensor structure proportional to $\delta_{ij}$; the latter
are eliminated by counterterms.
Thus the physical electric and magnetic polarization functions
become
\begin{align}
\itDelta \Pi_E^{ {\rm phys} } 
= &P^E_{\mu \nu} 
\left( \itDelta \bPi_{\mu \nu} 
+ \delta_{ {\rm c} } \bPi_{\mu \nu}^{ {\rm gaps} } \right) 
\nonumber\\
&= \itDelta \bPi_E 
- \frac{\, k_0^2 \,}{k^2} \, C_{ {\rm gaps} } (\vk^2)  \,,
\nonumber \\
\itDelta \Pi_M^{ {\rm phys} } 
=& \frac{1}{\, 2\,} P^M_{\mu \nu} 
\left( \itDelta \bPi_{\mu \nu} 
+ \delta_{ {\rm c} } \bPi_{\mu \nu}^{ {\rm gaps} } \right)
\nonumber\\ &= \itDelta \bPi_M - C_{ {\rm gaps} } (\vk^2) \,.
\label{physpol}
\end{align}
By construction 
$k_\mu \itDelta \Pi_{\mu \nu}^{ {\rm phys} } =0$.
While naive regularization 
with spatial cutoff does not affect the electric mass defined at $k_0=0$,
the magnetic mass requires modification.
Substituting the explicit form of $C_{ {\rm gaps} }$ at $k_0=0$
(simply $\itDelta \bPi_{ Ls} $),
we have
\beq
\itDelta \Pi_M^{ {\rm phys} } (k \rightarrow 0)
= \itDelta \bPi_M (k\rightarrow 0)
- \itDelta \bPi_{Ls} (k\rightarrow 0)  \,.
\eeq
In particular, 
 since we proved in Eq. (\ref{theorem}) that for the singlet or normal phases
$\itDelta \bPi_M (k\rightarrow 0) = \itDelta \bPi_L (k\rightarrow 0)$, 
the above expression shows that 
the magnetic mass must disappear, as stated earlier.

Finally let us return to the discussions about
the tensor structure of the counterterms in the Higgs phase,
$\delta_{ij}$ or $k_i k_j/\vk^2$.
As shown in the Eq.~(\ref{footnote}),
the vertex structure for the constant gap 
must be of the form,
\beq
{\bf \Gamma}_{\mu a}^{\calA} (q_+,q_-) 
\sim 2 \,\frac{\, k_0 g_{\mu 0} + v^2 k_j g_{\mu j} \,}
{\, k_0^2 + v^2 \vk^2 \,} \left[
\begin{matrix}
~ 0 & \Delta_c \gamma_5~\\
~ \gamma_5 \Delta  ~ & 0 ~
\end{matrix}
\right] \,.\nonumber\\
~~~~~[{\rm U(1)~Higgs~phase}]\nonumber\\
\label{footnote2}
\eeq
On the other hand, the present gauge-variant contributions
are functions of $\vk^2$, and not $k_0^2$.
Again we conclude that the counterterm
is $\sim\delta_{ij}$, and
we can continue to use Eq.~(\ref{physpol}),
although the actual terms in $\bPi_M$ and $\bPi_{Ls}$
are very different in the normal and singlet phases.

%%%%%%%%%%%%%%%%%%%%%%%%%%%%%%%%%%%%%%
\section{Numerical results}
\label{numerical}
%%%%%%%%%%%%%%%%%%%%%%%%%%%%%%%%%%%%%%

In this section we numerically evaluate the electric and magnetic 
masses for the normal, U(1)$_{ {\rm em} }$ Higg,
and singlet phases.
Results are presented for the (subtracted)
physical polarization functions, $\itDelta \Pi_{ {\rm phys} }$, including
corrections from vertices and counterterms.
We take
the effective quark mass in the vacuum subtraction to be $M_\chi = 300$ MeV, unless otherwise stated.
In most cases we present results normalized by the
square of the electric mass in normal phase divided by $g_s^2$,
\beq
\frac{\, m^2_{E,{\rm normal} } (k) \,}{\, g_s^2 (k) \,} 
\bigg|^{ {\rm one-loop}}_{k_0=0,\vk\rightarrow 0}
&= \Pi_E^{ {\rm normal} } (k_0=0,\vk\rightarrow 0) \nonumber\\
&= \Nf \frac{\, \mu^2 \,}{\, \pi^2 \,} \,.
\eeq
(The reason for dividing by $g_s^2$
is that in comparing  the vacuum and medium gluon polarization functions
both have an overall factor $g_s^2$.
At large $\alpha_s$,
not only medium masses but also vacuum gluon contributions should be
regarded as large quantities; thus, it is more natural for the purposes of comparison
to consider $m_E^2/g_s^2$
instead of $m_E^2$ itself. In fact,
in the present one-loop calculations,
the only place where large $\alpha_s$ enters
is in the sizes of the gaps.)

We present all results for the constant gaps, of various magnitudes,
to examine the impact of the size of the gaps.
We do not give results here for 
momentum-dependent gaps, since 
they require using improved vertices
whose explicit expressions are given only 
for the infrared limit in this paper.
The extension to finite momenta, which
requires explicit solutions of the
vertex functions, is deferred to a future paper.

\begin{figure}[t]
\vspace{0.0cm}
\begin{center}
\scalebox{0.5}[0.5] {
  \includegraphics[scale=0.45]{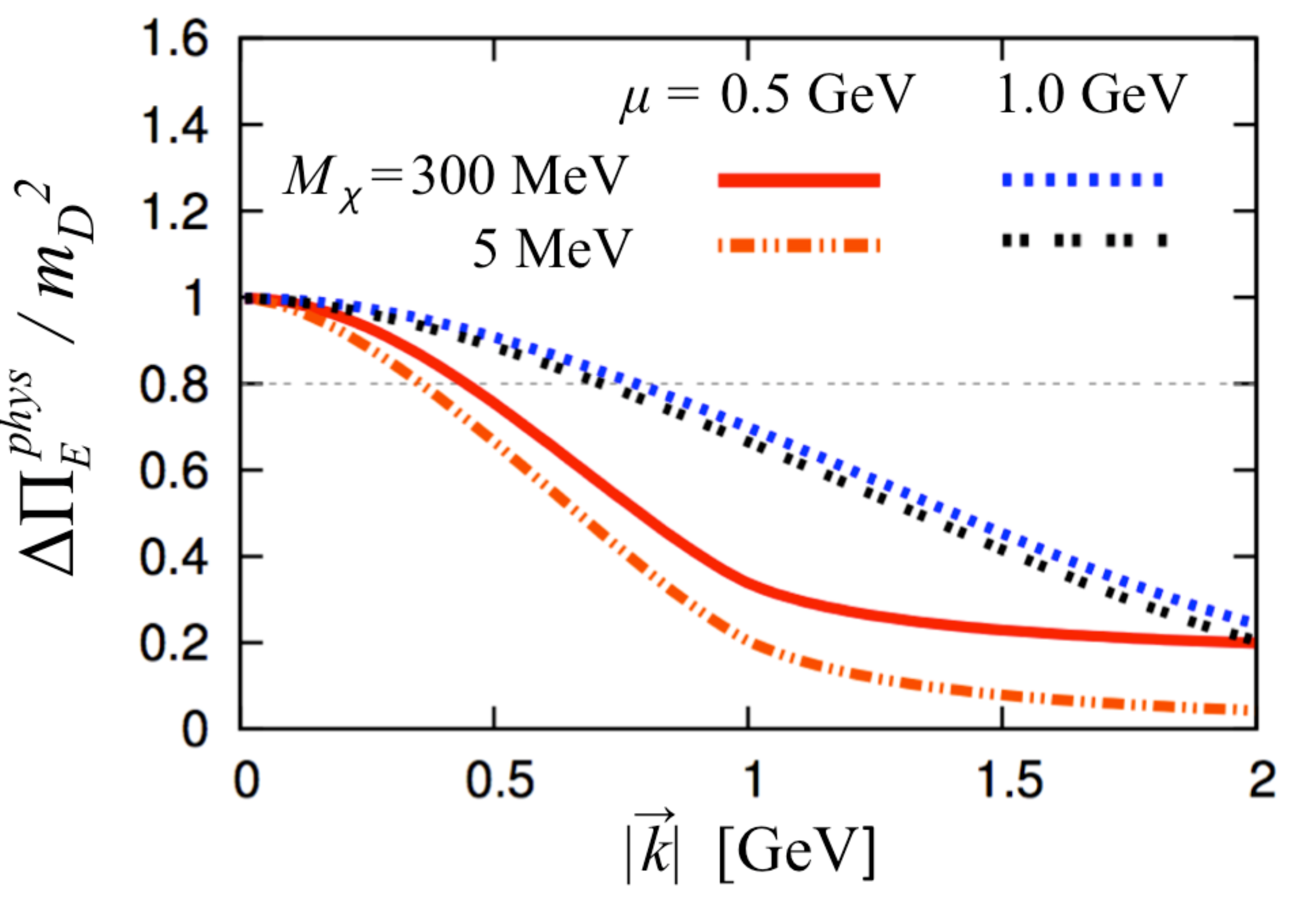} 
} \hspace{0.5cm}
\scalebox{0.5}[0.5] {
  \includegraphics[scale=0.445]{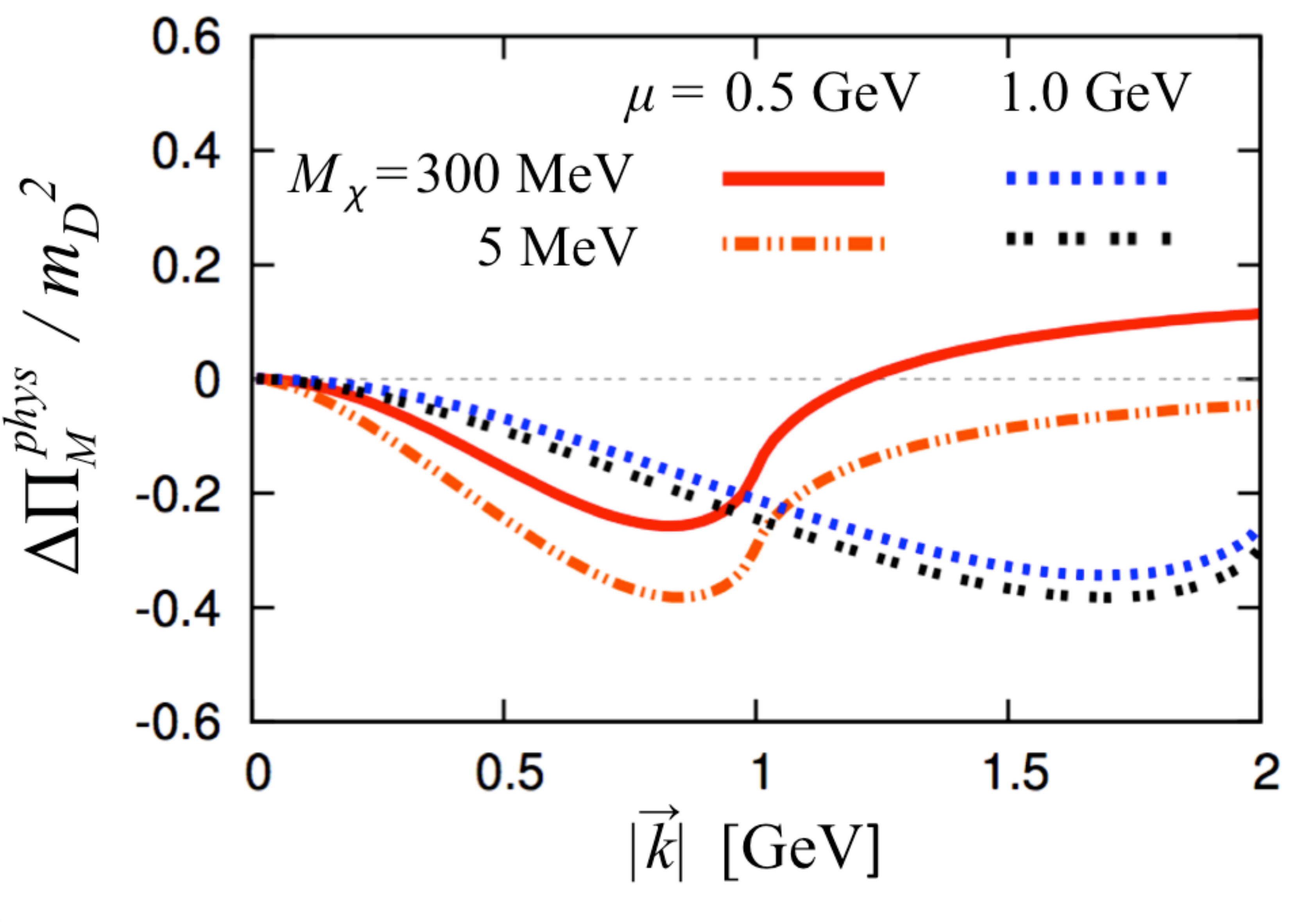} 
}
\end{center}
\vspace{-0.8cm}
\caption{Normal-phase 
static polarization functions 
$\itDelta \Pi_{ {\rm phys} } (k_0=0, \vk) 
= (\bPi - \bPi_{ {\rm vac} } )_{{\rm phys} }$
for $\mu =0.5$ and $1.0$ GeV,
normalized by the square of the Debye mass,
$m_D^2$.
We compare the vacuum subtractions for
different effective quark masses,
$M_\chi = 300$ and $m_{u,d}=5$ MeV.  
The upper panel is for the electric sector, where we draw a line in the upper panel for $\lqcd^2/m_D^2 \simeq 0.8$
at $\mu=0.5$ GeV
for comparison with the nonperturbative scale.
The IR contributions are larger than
$\sim \lqcd^2$.
The magnetic sector is shown in the lower panel.
The negative region appears mainly because of 
the particle-hole contributions 
(see also Fig.~\ref{fig:difnormalEM}).
}
\label{fig:normalEM}
\end{figure}
\begin{figure}[tb]
\vspace{0.0cm}
\begin{center}
\scalebox{0.5}[0.5] {
  \includegraphics[scale=0.45]{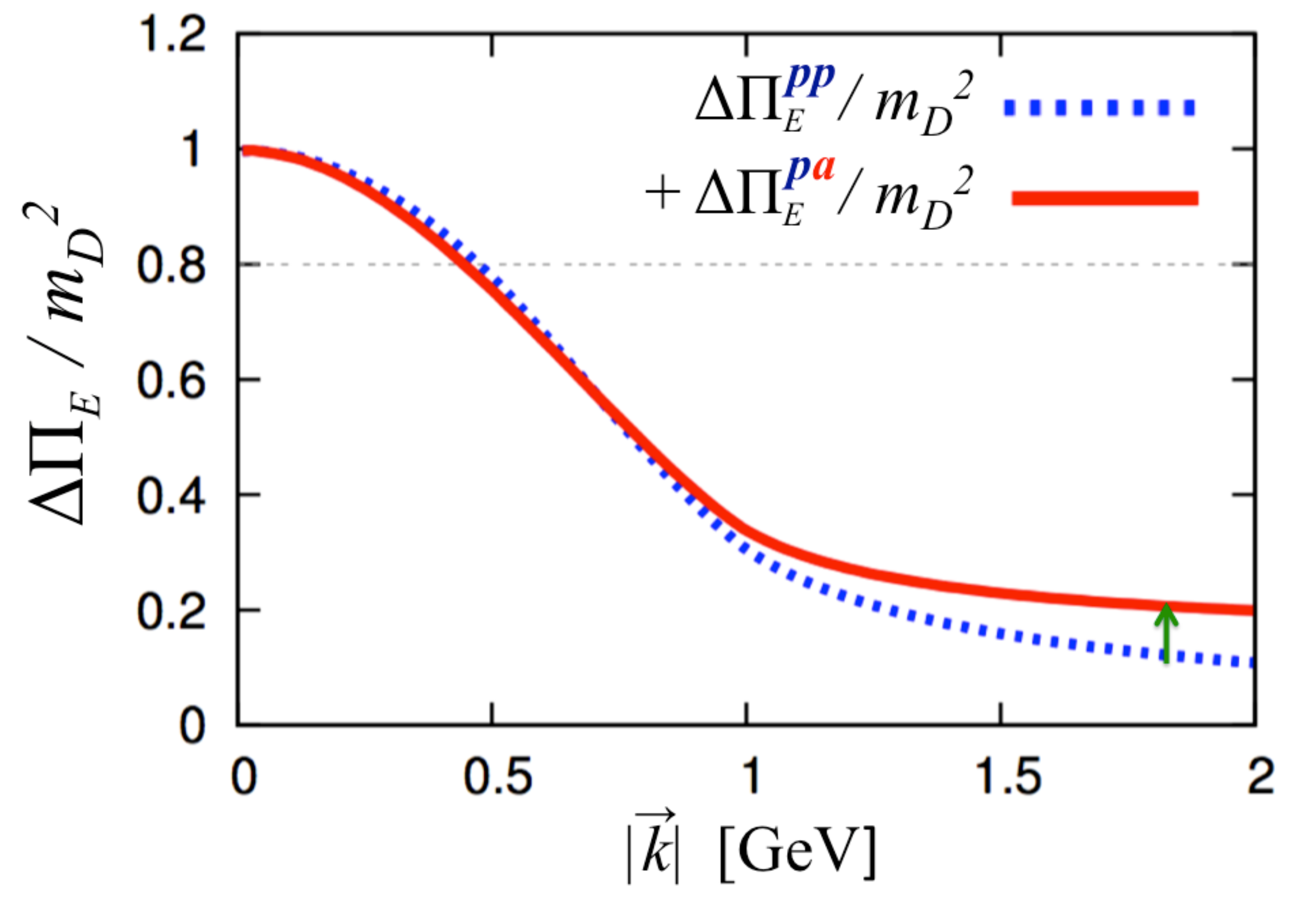} 
} \hspace{0.5cm}
\scalebox{0.5}[0.5] {
  \includegraphics[scale=0.445]{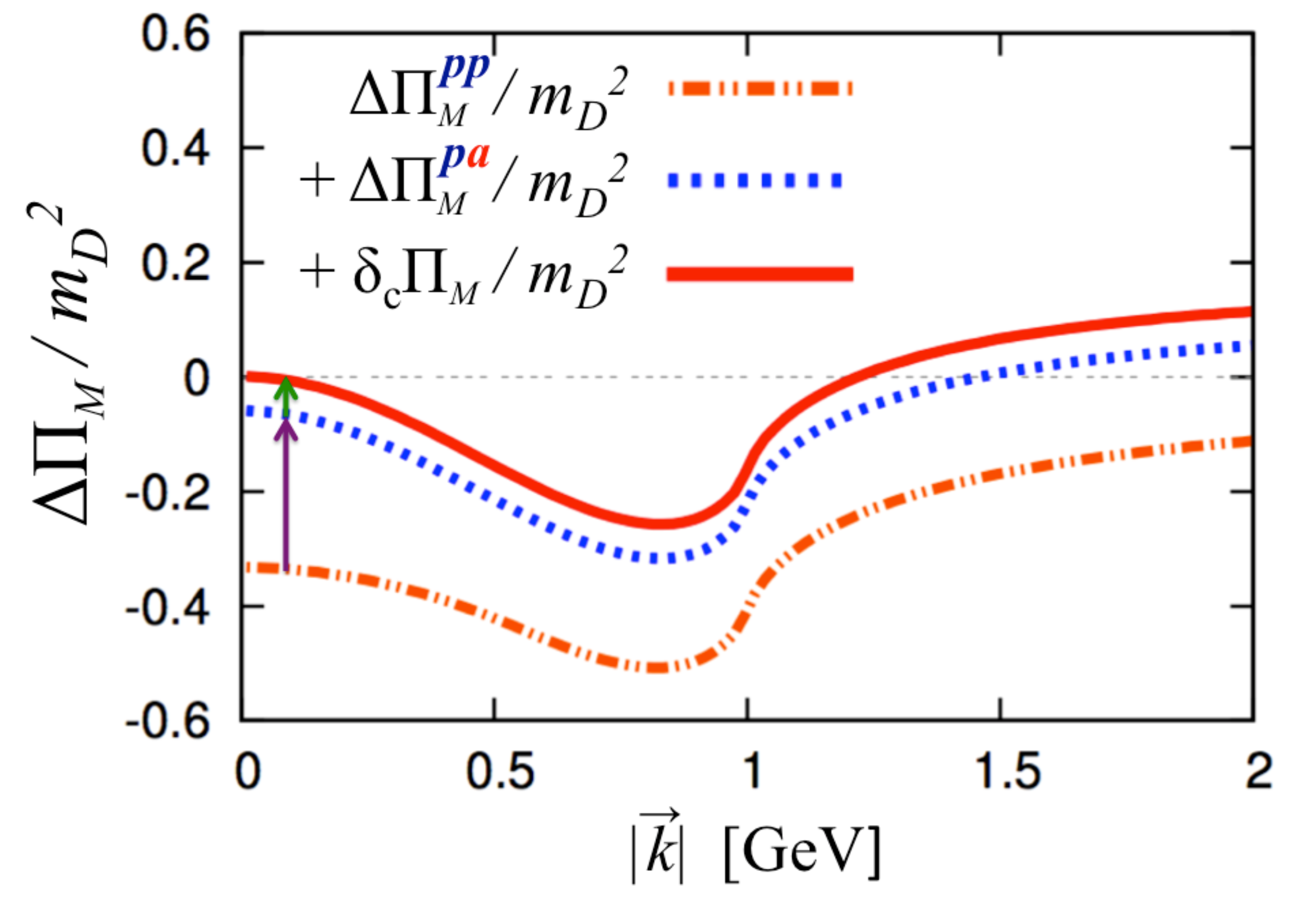} 
}
\end{center}
\vspace{-0.8cm}
\caption{The normal-phase results at $\mu =0.5$ GeV.
We sequently add  the 
particle-hole (pp), particle-antiparticle (pa)
and finally the counterterm contributions to recover the gauge invariance.
The upper panel shows the electric sector, where in
the static limit,
there are no gauge-variant contributions.
The magnetic sector is shown in the lower panel.  Here the particle-hole and particle-antiparticle
contributions in gauge-variant computations
almost cancel out;
the remainder is precisely cancelled out
by subtracting the gauge-variant contributions
that emerge from our regularization scheme. 
} 
\label{fig:difnormalEM}
\end{figure}
%
%%%%%%%%%%%%%%%%%%%%%%%%%%%%%%%%%%%%%%
\subsection{Normal phase}
\label{normalphase}
%%%%%%%%%%%%%%%%%%%%%%%%%%%%%%%%%%%%%%

We start with numerical results for the normal phase.
Although this situation has been studied previously,
the current quark mass was typically used in the chiral-symmetric vacuum. 
Our main purpose here is to examine
the effects of the vacuum subtractions
for the dynamical mass $M_\chi$ 
and the current quark mass.
For the latter we set $m_{u,d} = 5$ MeV.
Another reason for revisiting the normal-phase results
is to use them as a reference point to see the
characteristic features of the condensed phases.

In Fig.~\ref{fig:normalEM},
we plot $\itDelta \Pi_{E,M}^{ {\rm phys} } (k)$
in the static limit, $k_0=0$,
for different values of quark effective mass.
At small $\vk$, different vacuum subtractions
do not significantly affect the results that include proper treatments 
of the gauge-variant contributions.
(However, before taking care of the artifacts, one finds qualitative differences
in the magnetic sector; see below.)
However, their asymptotic behaviors at large $\vk$
are different.
The contributions from the renormalized
vacuum polarization function $\Pi^R_{ {\rm vac} }$
grow like
$\Pi^R_{ {\rm vac} } \sim \vk^2 \ln (\vk^2/\lambda_R^2)$.
After one adds $\itDelta \Pi_{E,M}^{ {\rm phys} } (k)$
and $\Pi^R_{ {\rm vac} }$ to find $\Pi^R$ in the medium,
$\Pi^R_{ {\rm vac} }$ becomes the dominant contribution. In both the IR and UV regions,
different vacuum subtractions do not produce significantly different results
{\it after} one takes care of the gauge-variant contributions.
This statement becomes more solid for
larger chemical potentials.

 At $\mu =0.5$ GeV,
the size of the electric screening
exceeds $\sim \lqcd^2$ in the IR region,
suggesting that electric gluons are well screened.
On the other hand, magnetic gluons
are protected from screening in the static limit,
unless $\alpha_s(k)$ in the infrared
shows significant 
enhancement \footnote{The behavior of the
quark-gluon vertex in vacuum can be
quite different for different choices of gauge
fixing conditions.
This discussion is beyond our scope in this work.}.
The dominant screening effect occurs at finite
frequencies (Landau damping);
overall it is large, 
and behaves like $\sim m_E^2 k_0/|\vk|$.

In Fig.~\ref{fig:difnormalEM},
we compare the roles of 
particle-hole and particle-antiparticle contributions
at $\mu =0.5$ GeV.
In the electric sector, the contributions are fairly dominated
by particle-hole contributions, for 
purely kinematic reasons, as we emphasized
in Sec.~\ref{kinematic factors}.
In the electric sector
gauge-variant artifacts are absent in the static limit.

On the other hand, in the magnetic sector,
the particle-hole excitations give {\it negative} contributions,
which are well cancelled by {\it positive} particle-antiparticle 
contributions.
The surviving contribution is 
just the gauge-variant artifacts introduced by
our regularization schemes with a spatial cutoff.
The size of the artifacts are $\sim 10$\% of the total
for $M_\chi =300$ MeV.
Had we set the mass terms 
in the normal quark matter
and in vacuum to be equal,
this gauge-variant contribution would be absent
from the very beginning, as was found in
the conventional hard-loop approximation.

%%%%%%%%%%%%%%%%%%%%%%%%%%%%%%%%%%%%%%
\subsection{Higgs phase}
\label{Higgsphase}
%%%%%%%%%%%%%%%%%%%%%%%%%%%%%%%%%%%%%%

%
\begin{figure}[tb]
\vspace{0.0cm}
\begin{center}
\scalebox{0.5}[0.5] {
  \includegraphics[scale=0.45]{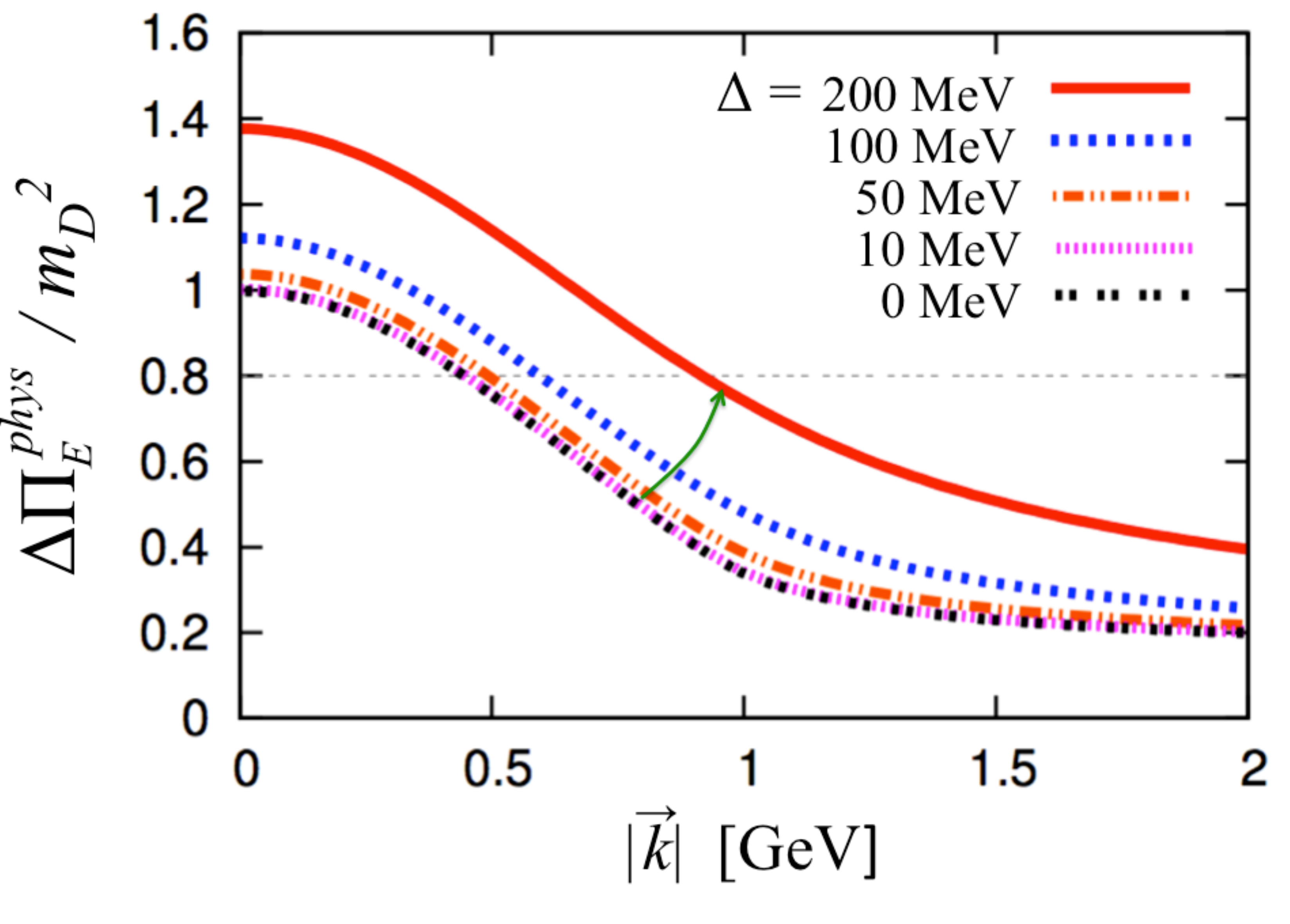} 
} \hspace{0.5cm}
\scalebox{0.5}[0.5] {
  \includegraphics[scale=0.445]{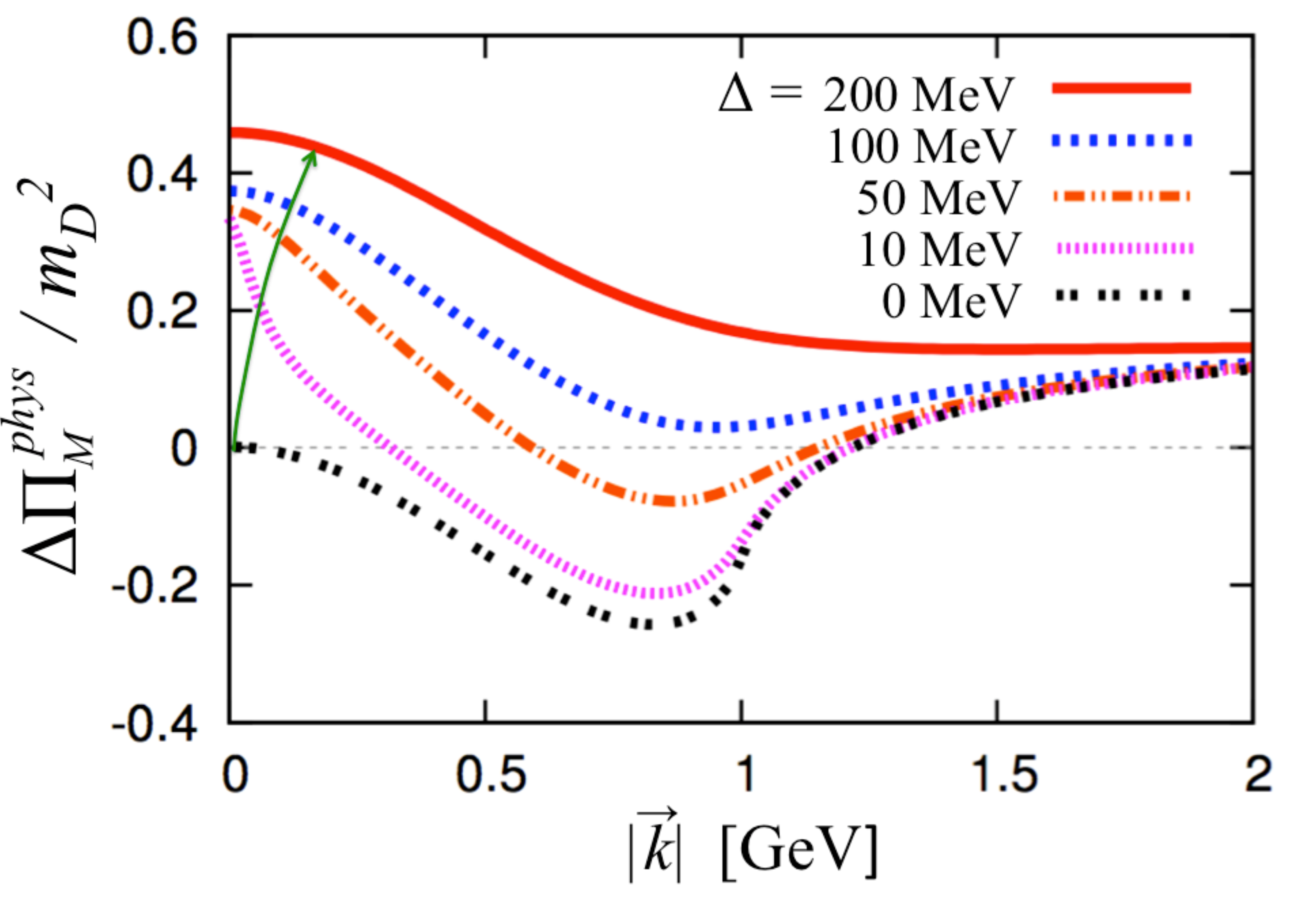} 
}
\end{center}
\vspace{-0.8cm}
\caption{The static polarization function $\itDelta \Pi_{ {\rm phys} } (k)$  in the Higgs phase
at $\mu = 0.5$ GeV.
We set $\Delta_\rma =0$ and vary $\Delta_\rmd$ from 10 to 200 MeV.
The normal-phase results and
the $\lqcd^2/m_D^2 \simeq 0.8$ line
are also plotted as guides.
The electric sector is shown in the upper panel.
The square of the electric mass, $m_E^2$,
in the Higgs phase 
is larger than that in the normal phase
by a factor $\sim (1+O(1) \Delta^2/\mu^2)$.
In the magnetic sector shown in the lower panel,
the infrared limit gives the Meissner mass, $m_M^2$.
As we reduce the size of the gap, 
$m_M^2$ approaches $m_E^2/3$.
At momenta beyond $\sim \Delta$,
the results start to approach those of the normal phase.
}
\label{fig:HiggsEM}
\end{figure}
\begin{figure}[ht]
\vspace{0.0cm}
\begin{center}
\scalebox{0.5}[0.5] {
  \includegraphics[scale=0.45]{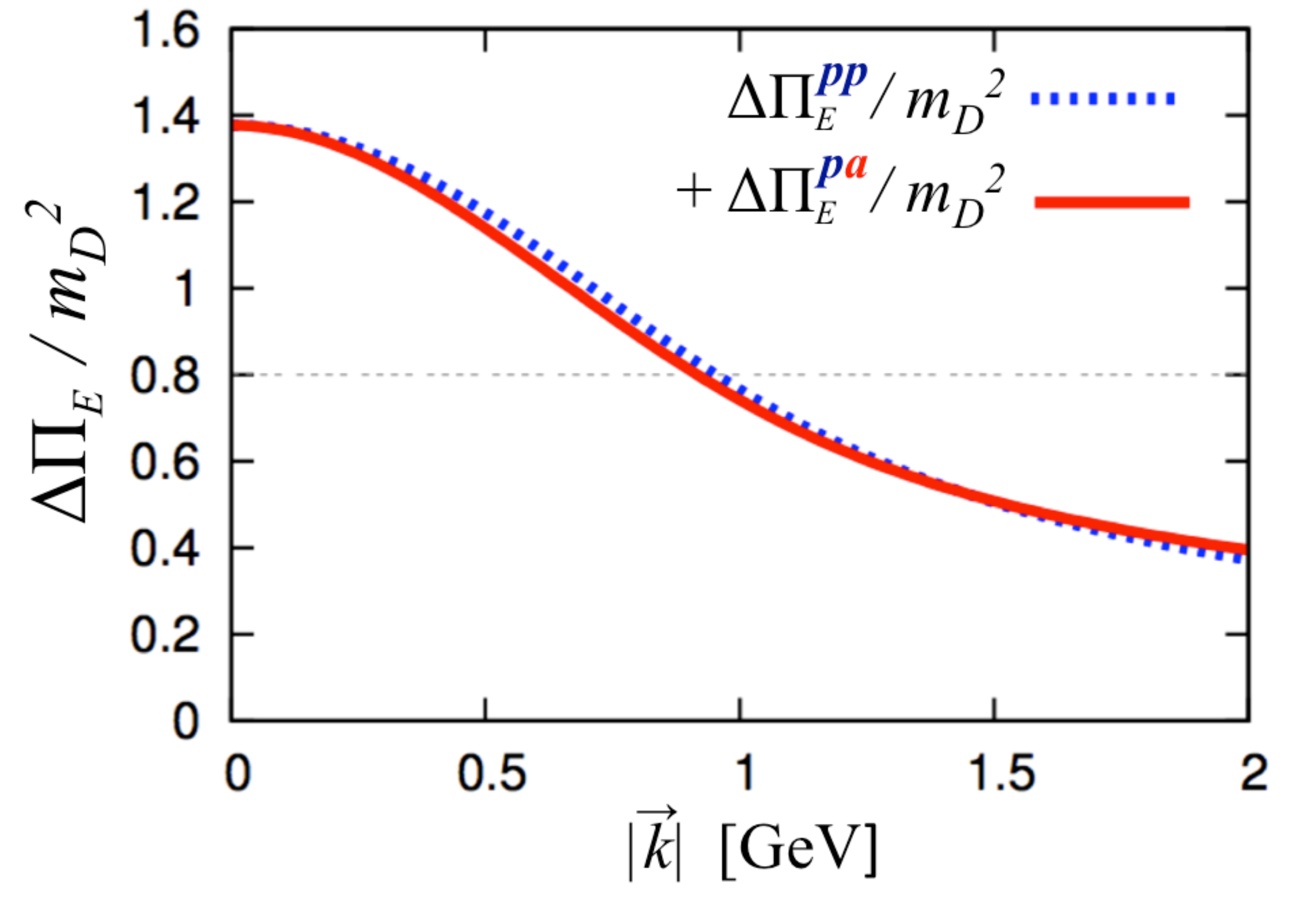} 
} \hspace{0.5cm}
\scalebox{0.5}[0.5] {
  \includegraphics[scale=0.45]{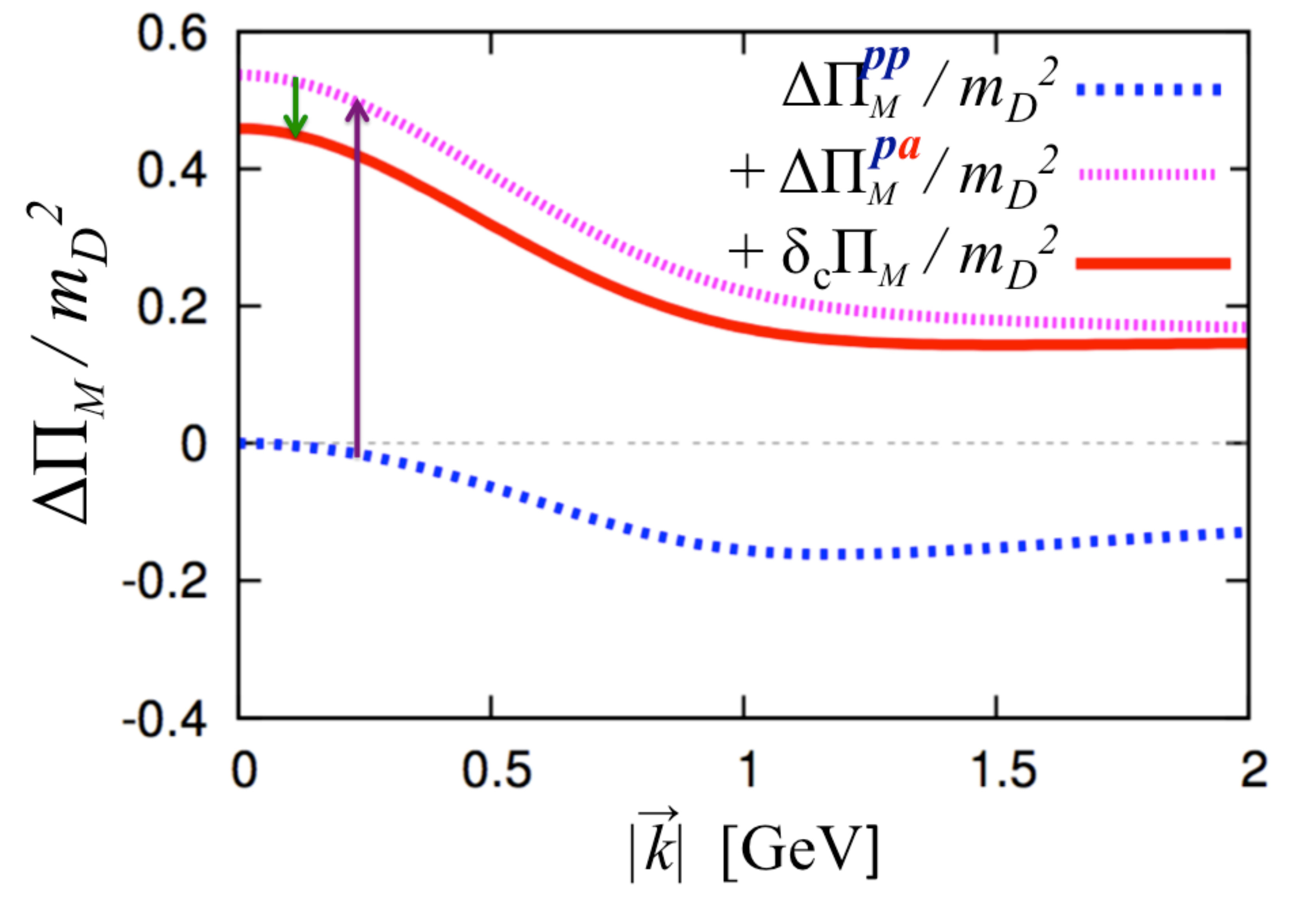} 
}
\end{center}
\vspace{-0.8cm}
\caption{The Higgs-phase results at $\mu =0.5$ GeV 
for $\Delta_\rmd = 200$ and $\Delta_\rma = 0$ MeV.
 As in Fig.~\ref{fig:difnormalEM}, we sequently add the contributions.
In the electric sector (upper panel)
the particle-antiparticle contributions are small, while in
the magnetic sector (lower panel)
the particle-hole contribution vanishes as $\vk \to 0$. The dominant contribution is 
that from the particles-antiparticle pairs.
Subtraction of the gauge-variant regularization artifacts gives
$\sim 10$ \% reduction.
} 
\label{fig:difHiggsEM}
\end{figure}

In Fig.~\ref{fig:HiggsEM},
we plot $\itDelta \Pi_{E,M}^{ {\rm phys} } (k)$
in the static limit, $k_0=0$,
with $\mu = 0.5$ GeV.
We focus on the gap near the Fermi surface,
setting the antiparticle gap $\Delta_\rma$ to zero, and letting
 $\Delta_\rmd$ = 10, 50, 100, and 200 MeV. 

The masses of the electric gluons are enhanced compared to their 
normal-phase values.
This enhancement can be understood as follows.
In a U(1)$_{ {\rm em} }$ superconductor,
the photon correlator in the infrared limit
is directly related to the correlator of the number density,
and the latter is related to the derivatives of the
thermodynamic pressure $P$ with respect to the number density $n$:
\beq
\frac{m_E^2}{g_s^2} = \frac{ \partial n }{\, \partial \mu \,}
= \frac{ \partial^2 P }{\partial \mu^2 },
\eeq
Since
the pressure is maximized in the ground state, the gap-dependent
terms in the Higgs phase increase the pressure, as
\beq
P_{ {\rm Higgs} } = c_{0} \mu^4 + c_2 \mu^2 \Delta^2 + \cdots \,,
\eeq
with $c_2>0$. 
Thus the electric masses in the Higgs and normal phases
are related by
\beq
\frac{ m^2_{ E, {\rm Higgs} } }{\, m^2_{E, {\rm normal}} \,}
\simeq 1 + \frac{c_2}{c_0} \frac{\, \Delta^2 \,}{\mu^2} \,.
\eeq
This tendency can be seen in Fig. \ref{fig:HiggsEM}.
In fact, at small $\Delta$ or large $\mu$,
the ratio quickly approaches $1$,
recovering the weak-coupling results.

In the magnetic sector,  the gaps do not strongly affect 
the overall magnitude of the magnetic mass;
rather, they substantially affect the size of domains
in which the polarization function differs from that in the normal phase.
In a weak-coupling computation of the gap,
the size of the IR domain where the gap plays a role is tiny, 
and the structure of the polarization in the magnetic sector depends on 
Landau damping without a Meissner mass, i.e., magnetic screening is 
negligible.
On the other hand, in strongly coupling treatments with a large gap,
the IR behavior of the magnetic sector  
is governed by the Meissner mass
instead of Landau damping, which is suppressed by the phase space.
The effects of the gap thus vary considerably with density.  
A detailed calculation of these
effects remains an interesting problem.

In Fig.~\ref{fig:difHiggsEM}, 
we compare the various  particle-hole, etc., contributions for
$(\Delta_\rmd,\Delta_\rma)=(200,0)$ MeV and $\mu =0.5$ GeV.
The particle-hole
contributions saturate the electric sector.
On the other hand,
in the magnetic sector,  the Higgs and normal phases are significantly different.
In the former, particle-hole contributions
precisely vanish, and the positive particle-antiparticle
contributions dominate the polarization functions.

Note that the gauge-variant contributions are $\sim 10$\%, 
a consequence
of the large gap, $\Delta_\rmd =200$ MeV.
For a small gap in a weak-coupling calculation,
the gauge-variant contributions are a quantitatively negligible
fraction of the total.

%%%%%%%%%%%%%%%%%%%%%%%%%%%%%%%%%%%%%%
\subsection{Singlet phase}
\label{singletphase}
%%%%%%%%%%%%%%%%%%%%%%%%%%%%%%%%%%%%%%

%
\begin{figure}[tb]
\vspace{0.0cm}
\begin{center}
\scalebox{0.5}[0.5] {
  \includegraphics[scale=0.45]{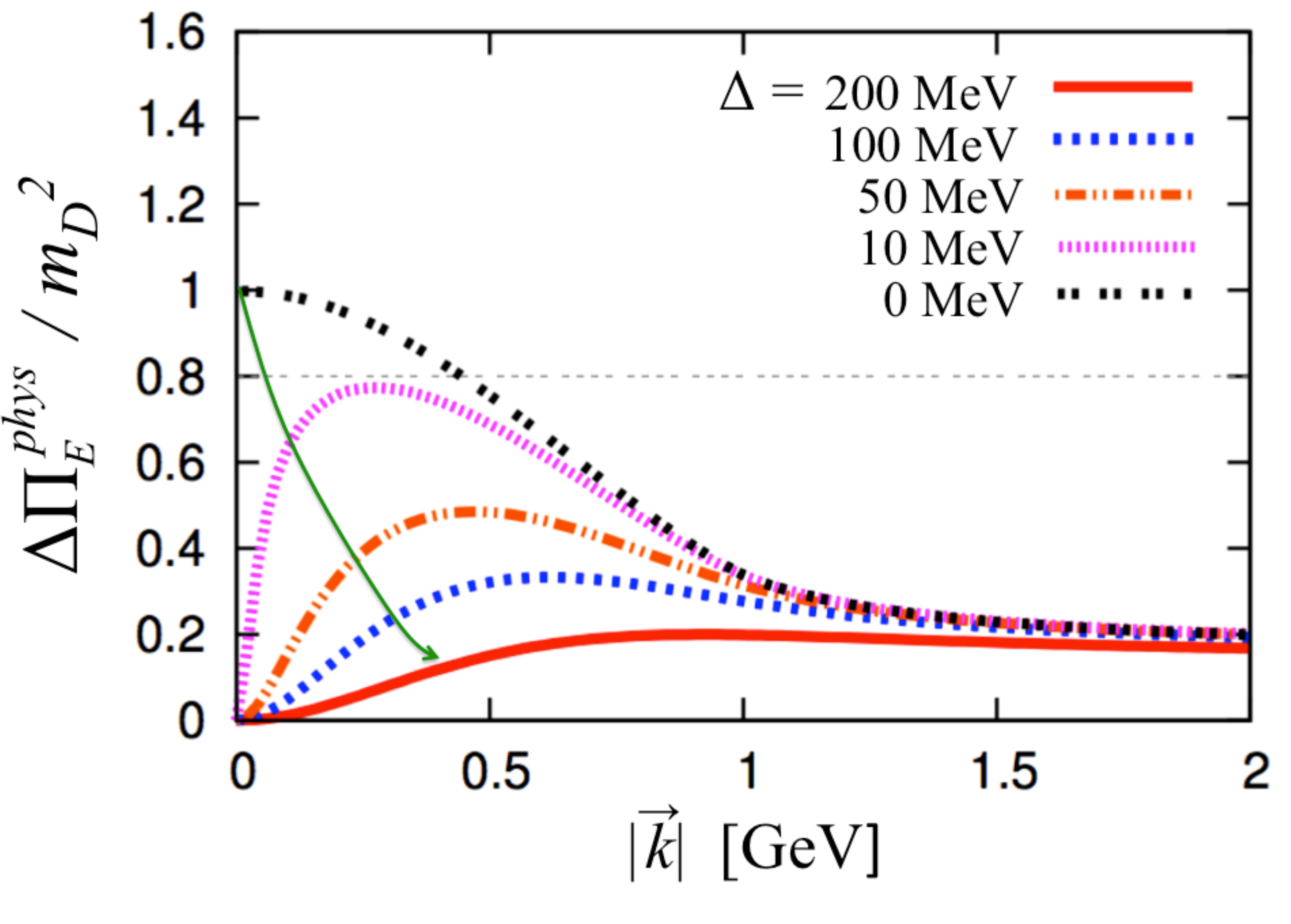} 
} \hspace{0.5cm}
\scalebox{0.5}[0.5] {
  \includegraphics[scale=0.45]{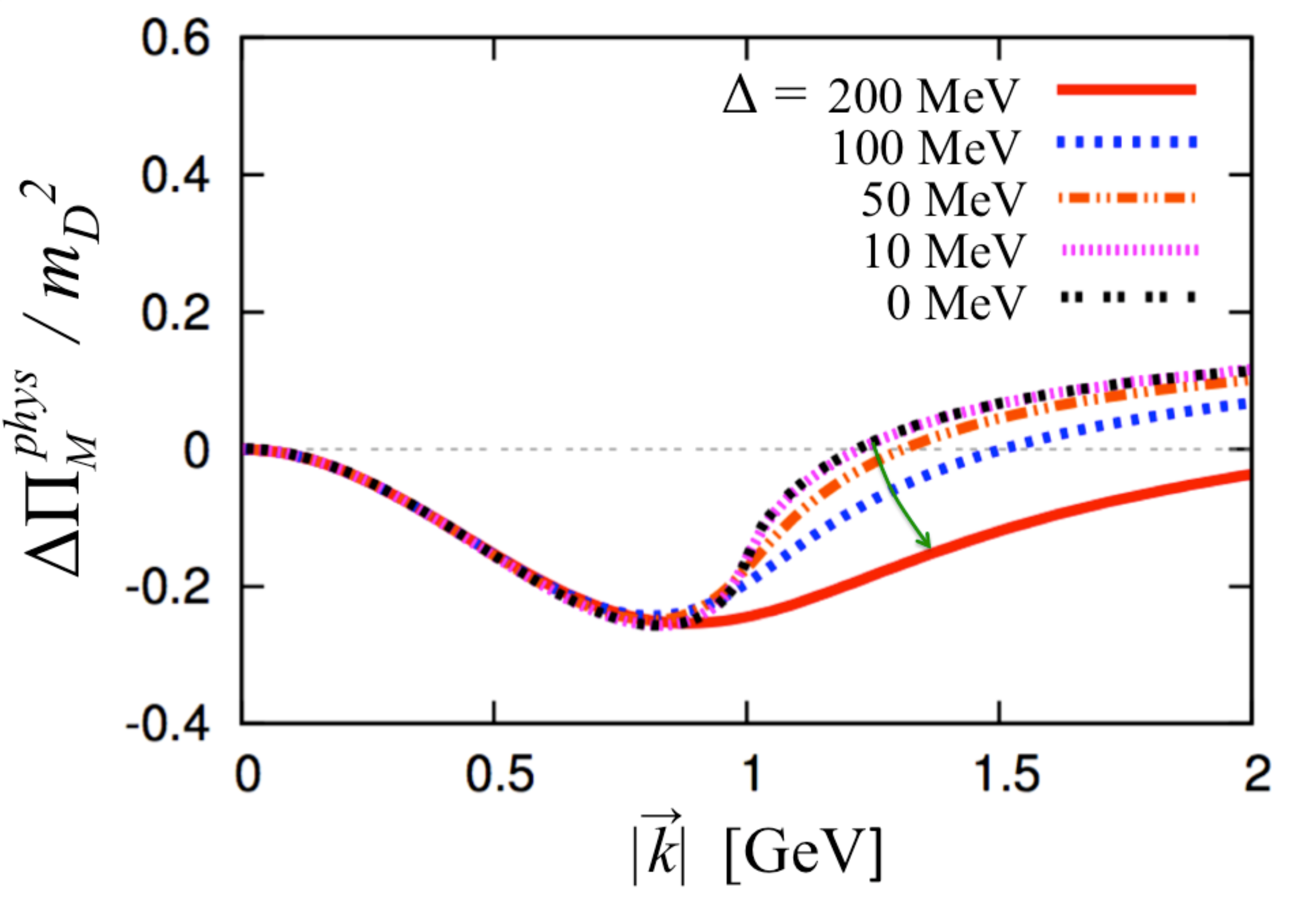} 
}
\end{center}
\vspace{-0.8cm}
\caption{The singlet-phase static polarization functions 
$\itDelta \Pi_{ {\rm phys} } (k)$.
As before, we set $\mu = 0.5$ GeV,
$\Delta_\rma =0$,
and vary the size of $\Delta_\rmd$ from 10 to 200 MeV.
In the electric sector (upper panel)
the infrared contributions are substantially 
suppressed compared to the normal phase,
and well below $\lqcd^2$.
The results for the infrared region in the magnetic sector (lower panel) 
are generally quite similar to those in the normal-phase results,
although around $|\vk| \sim 2p_F \sim 2\mu$
differences start to appear.
}
\label{fig:SingletEM}
\end{figure}

For the singlet phase we take the same
parameter set as for the Higgs phase.
Figure~\ref{fig:SingletEM} shows the behavior of the polarization functions.
The main differences from the normal and Higgs phases
can be seen in the electric sector.
The IR contributions are vanishing;
in particular, for $\Delta_\rmd =200$ MeV,
the electric contribution is well below that in
the normal phase.
The quark color density is much stiffer
against color perturbations than in the normal phase,
which implies that electric gluons in the IR region 
are unaffected by screening,
unless $\alpha_s(k)$ is significantly enhanced in the infrared.
The size of the unscreened domain 
is characterized by the size of the gap, shrinking
as $\Delta_\rmd$ decreases.

The infrared behavior in the magnetic sector,
is quite similar to that in the normal phase.   
While its behavior in
the UV is different, it has little
quantitative impact on the total, where
vacuum contributions growing like $\sim k^2$ become large.

\begin{figure}[tb]
\vspace{0.0cm}
\begin{center}
\scalebox{0.5}[0.5] {
  \includegraphics[scale=0.45]{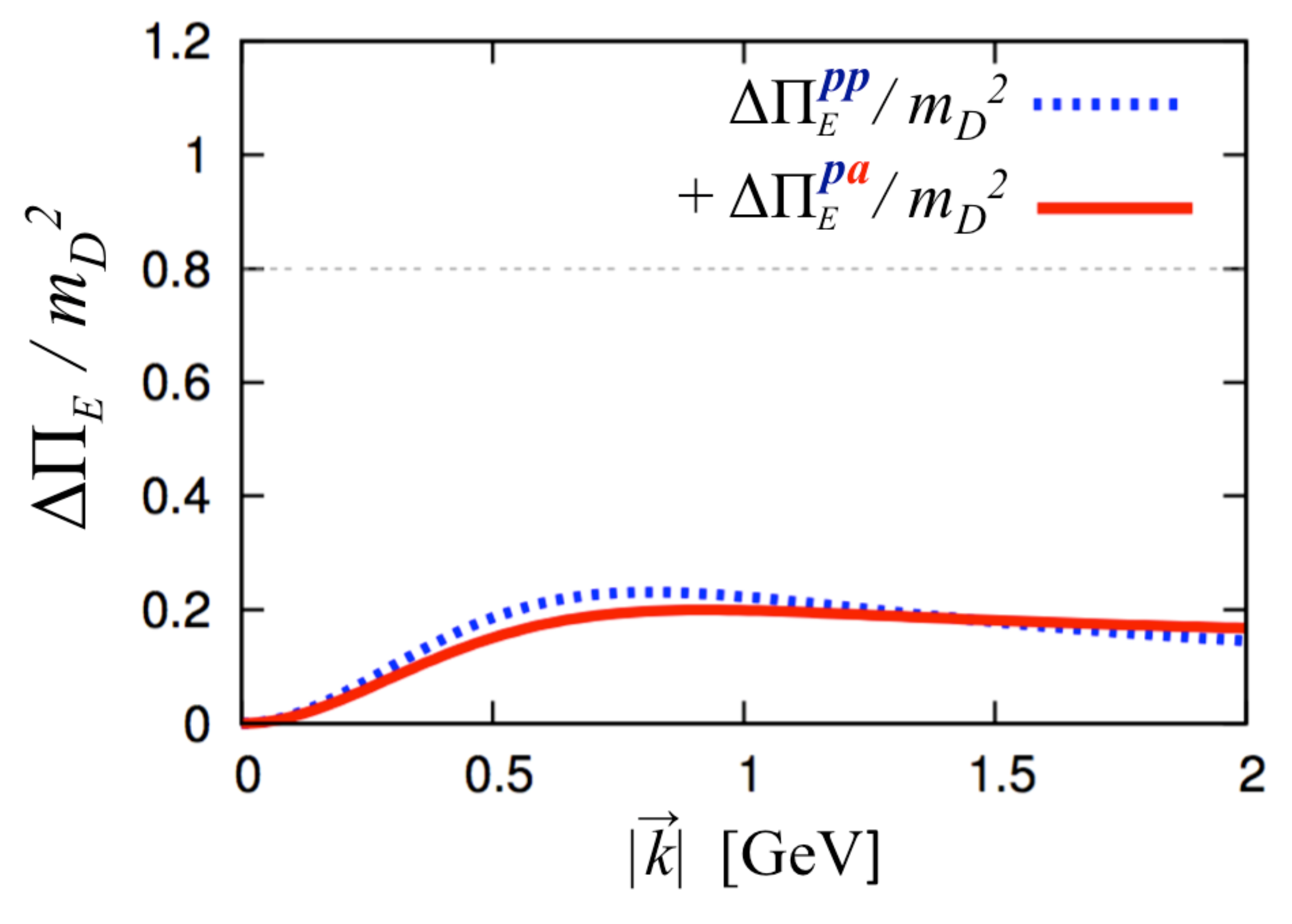} 
} \hspace{1.0cm}
\scalebox{0.5}[0.5] {
  \includegraphics[scale=0.45]{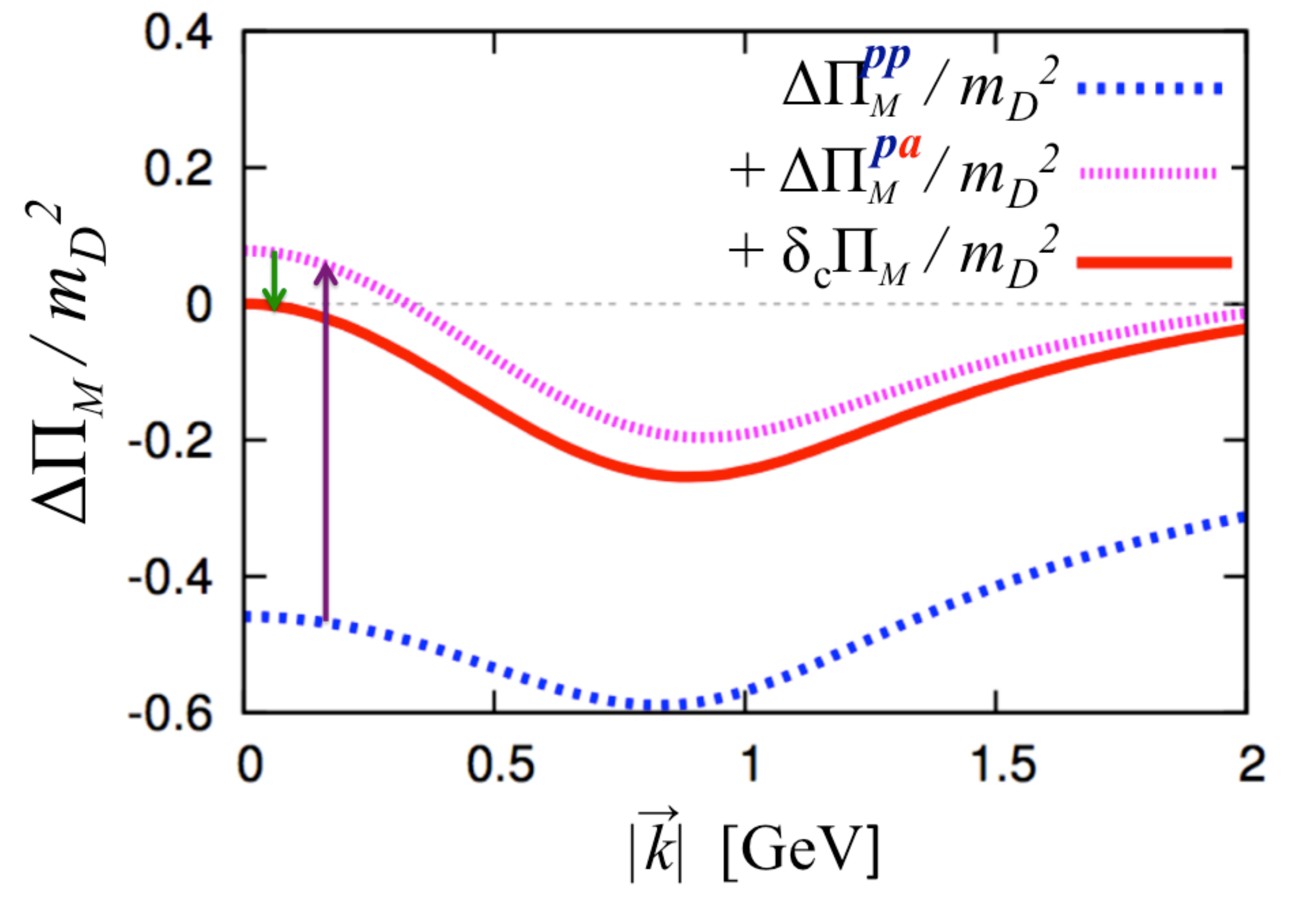} 
}
\end{center}
\vspace{-0.6cm}
\caption{The singlet-phase results at $\mu =0.5$ GeV
for $\Delta_\rmd = 200$ MeV and $\Delta_\rma = 0$ MeV.
We sequentially add contributions as in Fig. \ref{fig:difnormalEM}.
In the electric sector (upper panel)
particle-hole contributions are dominant.
In the magnetic sector (lower panel) the particle-hole and particle-antiparticle
contributions tend to cancel out, as in the normal phase,
and the remaining contribution is precisely eliminated
by subtracting the gauge-variant terms.
} 
\label{fig:difSingletEM}
\end{figure}

Figure~\ref{fig:difSingletEM} also shows comparisons of the various contributions.
The electric sector is well dominated by 
particle-hole contributions; 
in the magnetic sector,
the situation is similar to that in the normal phase.
The particle-hole and particle-antiparticle contributions
almost cancel out,
and the remaining contributions are gauge-variant artifacts.  
We conclude that the in-medium gluons 
in the singlet phase behave 
in the infrared as vacuum gluons.

%
%%%%%%%%%%%%%%%%%%%%%%%%%%%%%%%%%%%%%%
\section{Summary and discussions}
\label{summary}
%%%%%%%%%%%%%%%%%%%%%%%%%%%%%%%%%%%%%%

In this paper we have compared color screening
in the normal, Higgs, and singlet phases. 
We studied the singlet phase using the example of two-color QCD 
with a color-singlet condensates
near the Fermi surface.
The presence of the gap provides
qualitative differences among these phases.
In particular,
in the singlet phase 
both the electric and magnetic screening masses disappear,
implying that soft gluons are protected from 
medium effects, as long as the quark-gluon vertex is not singular in the infrared.

An obvious question concerns the three-color version of the
singlet phase.
With two colors, the natural singlet condensate
is a uniform diquark condensate.
In contrast, in three-color QCD the diquark is colored,
so we have to look for alternative condensates 
to carry over our arguments here.
The usual uniform chiral condensate 
formed by particle-antiparticle pairs is not
favored in the presence of a quark Fermi sea;
instead one might imagine a uniform particle-hole
condensate,
but the allowed phase space is too small to favor such a 
condensate.
In fact the usual gap equation for uniform chiral condensation 
at finite density automatically includes this possibility but does not
yield a nontrivial solution.

One possible candidate would be a nonuniform chiral condensate
made of particles and holes.
The structure of the gap equation is like that in the BCS one, 
and the size of the gap
can be enhanced by the quasi-low dimensionality 
near the Fermi surface.
If gluons remain strongly interacting at densities of interest,
the gap can be $\sim \lqcd$;
such a large gap can protect 
soft gluons from medium effects 
as discussed here, giving a self-consistent picture.
In this context, studies of the non-uniform chiral condensates
deserve further investigation.

Strong interactions of gluons---were they to remain 
up to $\mu \simeq 0.5$ GeV or larger---would justify a number of tacit assumptions in
frequently employed effective-model calculations, e.g., 
the Nambu-Jona-Lasinio model.
Effective models are usually formulated to describe
the hadron phenomenology,
incorporating gluon dynamics into a 
set of model parameters or particular forms of interactions, which can
in principle change
if the underlying gluon dynamics changes in the medium.
The presence of condensates
that forbid significant modifications in gluon dynamics would render such effective-model treatments consistent
at finite density.
Furthermore, the infrared protection of soft gluons
would leave the gluon condensate---which is related to the
QCD vacuum energy density---essentially unchanged.
In this picture, the quark matter equation of state would
not contain an additional constant term, e.g.,  a ``bag constant."

Another issue is the treatment of strange quarks.
It is generally assumed that the strange quarks do not play a role
until $\mu$ becomes close to the strange-quark 
effective mass, $\sim 500$ MeV.
But the origin of the effective quark mass---chiral symmetry breaking---
would disappear or significantly decrease
once soft gluons are strongly screened; then
a strange-quark Fermi sea would be formed
much earlier than one would expect with 
a phenomenological strange-quark constituent mass,
because the strange-quark current mass, $\sim 100$ MeV, is
well below the typical scale for quark matter formation, $\mu \sim$ $300-400$ MeV.
The role of such an early onset of strange quarks in reducing the stiffness 
of the quark matter has been explored in
Ref. \cite{pnjlphases}.

Finally we compare the present one-loop considerations
with lattice results for two-color QCD, which have studied the Landau-gauge gluon propagator 
in the presence of a
color-singlet diquark condensate \cite{Cotter:2012mb}.
The lattice results,
indicate that both
electric and magnetic gluons in the infrared region
are screened by medium effects, and look like gluons
in the Higgs 
phase \footnote{This description may
be misleading because a number of studies have 
indicated that
the confined and Higgs phases can be smoothly 
connected \cite{Fradkin:1978dv,Greensite:2003bk}.
Indications are based on gauge-invariant correlation functions
in which quarks and gluons are not separately discussed 
and all excitations are composite,
with no color charges.  
Such a connection between the two phases is obscured in gauge-fixed 
computations using quark and gluon propagators, as here.}.
These results are most likely nonperturbative---with, 
as  one expects, screening masses of order $gT$ or $g\mu$---and 
cannot be interpreted within our present perturbative framework. 
Although our results have quantitative ambiguities,
most of the present qualitative conclusions have been derived from 
considerations of phase-space restrictions introduced by the gaps.  
Thus we do not expect that the simple inclusion
of higher-order loops to resolve the difference with the lattice calculations.
Rather this discrepancy serves as a clue for deeper understanding
of nonperturbative gluon dynamics
that is not included in our computations.

%%%%%%%%%%%%%%%%%%%%%%%%%%
\section*{Acknowledgments}
%%%%%%%%%%%%%%%%%%%%%%%%%%
This research was supported in part by NSF Grants PHY09-69790 and PHY13-05891.  
GB wishes to thank the RIKEN iTHES project for partial support.

\appendix
\section{Derivation of the Ward-Takahashi identity}
\label{DerivationWTI}

In this appendix we use the standard path-integral formalism
to derive the generalized identity (\ref{WTI}) from which follows the 
Ward-Takahashi identities used in Sec.~\ref{trans}.
Consider 
\beq
\left\la \prod_{n=1}^N O_n(z_n) \right\ra 
= Z^{-1}\!\int \calD \Phi 
~ \prod_{n=1}^N O_n(z_n)
~\rme^{ - S[\Phi ] } \,,
\eeq
where we write compactly
$\calD \Phi 
= \calD \psi \calD \bar{\psi} \calD A
\cdots \,.$

To derive the needed identities we change the integration variables
for quarks, writing
$\psi'(x) = \left(1+\rmi \alpha_a (x) T_a \right) \psi (x)$.
This change of variables preserves the functional measure, and does
not affect the expectation values resulting from integration.  With the
corresponding 
change of variables for the charge-conjugated fields,
the Nambu-Gor'kov bases in the new and old
variables are related as
\beq
\Psi' (x) = \left(1+\rmi \alpha_a(x) R_a \right) \Psi(x),
\eeq
with $R$ given by Eq.~(\ref{R}).
The action, however, is not invariant under these changes of variables;
using the relations $[T_a,T_b]=\rmi f_{abc} T_c$,
$[T_a^T,T_b^T]= - \rmi f_{abc} T^T_c$,
and the resulting relation $[R_a,R_b]=\rmi f_{abc} R_c$, we find
the additional contribution
\beq
\delta \left( \bar{\Psi} \Slash{D} \Psi \right)
=
- \rmi \alpha_a(x) \left[\, 
\partial_\mu j^a_\mu 
- f_{abc} A_\mu^b \bar{\Psi} \gamma_\mu R_c \Psi 
\right]\,.
\eeq
Since the change of integration variables does not
affect the expectation value $\la O(z) \ra$,
the collection of terms linear in $\alpha$
must sum to zero.
Writing the change of $O(z)$ as
$\int_x \alpha_a (x) \delta^D (x-z) \delta^a O (z)$,
we find the identity:
\beq
\int \calD \Phi 
\left[  D^{ac}_\mu j^c_\mu (x)  \prod_{n=1}^N O_n(z_n) + \sum_{n=1}^N \delta^D (x-z_n) \times
\right. \nonumber\\
 \left.
\times\prod_{m=1}^{n-1} O_m(z_n) \,
\delta^a O (z_n)
\prod_{m=n+1}^{N} O_m (z_m)
\right]=0\,,
\eeq
from which we derive several relations.
For example, for $O=1$,  we find the quantum equation of motion,
$\la D^{ac}_\mu j^c_\mu (x) \ra=0$.
Setting $O (z_1) O(z_2) = \Psi(z_1) \bar{\Psi} (z_2)$,
we find Eq.(\ref{WTI}).

%%%%%%%%%%%%%%%%%%%%%%%%%%%%

%%%%%%%%%%%%%%%%%%%%%%%
\end{document}